\documentclass[a4paper,11pt]{article}
\pdfoutput=1 

\usepackage{jheppub} 

\usepackage[T1]{fontenc} 
\usepackage{xcolor}
\usepackage{makecell}
\usepackage{float}
\usepackage{multicol}
\usepackage{xspace}
\usepackage{booktabs}
\usepackage{xspace}
\usepackage[shortlabels]{enumitem}

\newcommand{\ome}{(1-\epsilon)}
\newcommand{\e}{\epsilon}

\newcommand{\modsq}[1]{\ensuremath{\left\vert #1 \right\vert^2}}
\newcommand{\calM}{\ensuremath{\mathcal{M}}}
\newcommand{\order}[1]{\ensuremath{\mathcal{O}}\left( #1 \right)\xspace}

\renewcommand{\NG}[1]{\textcolor{black}{NG: #1}}

\newcommand{\Dox}{\mathcal{D}_{0}(\xi)}
\newcommand{\Doxi}{\mathcal{D}_{0}(x_{i})}
\newcommand{\Doxii}{\mathcal{D}_{0}(x_{k})}
\newcommand{\Dlx}{\mathcal{D}_{1}(\xi)}
\newcommand{\Dlxi}{\mathcal{D}_{1}(x_{i})}
\newcommand{\Dlxii}{\mathcal{D}_{1}(x_{k})}

\newcommand{\NthreeLO}{N$^3$LO\xspace}

\newcommand{\Pqg}{P_{qg}}

\newcommand{\Pgg}{P_{gg}}
\newcommand{\Pqq}{P_{q\bar{q}}}

\newcommand{\PggS}{P_{gg}^\text{sub}}

\newcommand{\PSup}{\mathrm{\mathbf{S}}^\uparrow\xspace}
\newcommand{\PSdown}{\mathrm{\mathbf{S}}^{\downarrow}\xspace}

\newcommand{\PCdownFF}{\mathrm{\mathbf{C}}^{FF,\downarrow}\xspace}
\newcommand{\PCupFF}{\mathrm{\mathbf{C}}^{FF,\uparrow}\xspace}

\newcommand{\PCdownIF}{\mathrm{\mathbf{C}}^{IF,\downarrow}\xspace}
\newcommand{\PCupIF}{\mathrm{\mathbf{C}}^{IF,\uparrow}\xspace}

\newcommand{\PCdownII}{\mathrm{\mathbf{C}}^{II,\downarrow}\xspace}
\newcommand{\PCupII}{\mathrm{\mathbf{C}}^{II,\uparrow}\xspace}

\newcommand{\PPup}{\mathrm{\mathbf{P}}^\uparrow\xspace}
\newcommand{\PPdown}{\mathrm{\mathbf{P}}^\downarrow\xspace}

\newcommand{\X}{X_3^0}

\newcommand{\A}{A_3^0}

\newcommand{\D}{D_3^0}
\newcommand{\F}{F_3^0}
\newcommand{\E}{E_3^0}

\newcommand{\G}{G_3^0}

\newcommand{\oldant}{\, \mathrm{OLD}}

\newcommand{\Xold}{X_3^{0,\oldant}}

\newcommand{\calX}{{\cal X}_3^0}
\newcommand{\calXold}{{\cal X}_3^{0,\oldant}}

\newcommand{\calA}{{\cal A}_3^0}

\newcommand{\calD}{{\cal D}_3^0}

\newcommand{\calF}{{\cal F}_3^0}

\newcommand{\calE}{{\cal E}_3^{0}}

\newcommand{\calSsoft}{\mathcal{S}\kern-0.05em{s\kern-0.05em o\kern-0.05em f\kern-0.05em t}}
\newcommand{\calScol}{\mathcal{S}\kern-0.05em{c\kern-0.05em o\kern-0.05em l}}
\newcommand{\calDsoft}{\mathcal{D}\kern-0.05em{s\kern-0.05em o\kern-0.05em f\kern-0.05em t}}
\newcommand{\calTcol}{\mathcal{T}\kern-0.05em{c\kern-0.05em o\kern-0.05em l}}
\newcommand{\calDcol}{\mathcal{D}\kern-0.05em{c\kern-0.05em o\kern-0.05em l}}

\newcommand{\qp}{Q}
\newcommand{\qpbar}{\bar{Q}}

\newcommand{\zj}{z_j}
\newcommand{\omzj}{(1-z_j)}

\newcommand{\omxj}{(1-x_j)}

\newcommand{\sij}{s_{ij}}
\newcommand{\sjk}{s_{jk}}
\newcommand{\sik}{s_{ik}}

\newcommand{\sijk}{s_{ijk}}
\renewcommand{\xi}{x_{i}}
\newcommand{\xj}{x_{j}}
\newcommand{\xk}{x_{k}}

\renewcommand{\NG}[1]{\textcolor{black}{#1}}

\newcommand{\scaleIF}{s_{\widehat{I}K}}
\newcommand{\QQIF}{\left(\frac{-\scaleIF}{\mu^2}\right)^{-\e}}
\newcommand{\scaleII}{s_{\widehat{I}\widehat{K}}}
\newcommand{\QQII}{\left(\frac{\scaleII}{\mu^2}\right)^{-\e}}
\newcommand{\IP}{\textrm{IP}}
\newcommand{\IC}{\textrm{IC}}

\newcommand{\qA}{A_{3,q}^0}
\newcommand{\qD}{D_{3,q}^0}
\newcommand{\qE}{E_{3,q}^0}
\newcommand{\gD}{D_{3,g}^0}
\newcommand{\gF}{F_{3,g}^0}
\newcommand{\gG}{G_{3,g}^0}
\newcommand{\gtoqA}{A_{3,g \to q}^0}
\newcommand{\gtoqD}{D_{3,g \to q}^0}
\newcommand{\QtogE}{E_{3,\qp \to g}^0}
\newcommand{\QtogG}{G_{3,\qp \to g}^0}

\newcommand{\calqA}{{\cal A}_{3,q}^0}
\newcommand{\calqD}{{\cal D}_{3,q}^0}
\newcommand{\calqE}{{\cal E}_{3,q}^0}
\newcommand{\calgD}{{\cal D}_{3,g}^0}
\newcommand{\calgF}{{\cal F}_{3,g}^0}
\newcommand{\calgG}{{\cal G}_{3,g}^0}
\newcommand{\calgtoqA}{{\cal A}_{3,g \to q}^0}
\newcommand{\calgtoqD}{{\cal D}_{3,g \to q}^0}
\newcommand{\calQtogE}{{\cal E}_{3,\qp \to g}^0}
\newcommand{\calQtogG}{{\cal G}_{3,\qp \to g}^0}

\newcommand{\qAold}{A_{3,q}^{0,\oldant}}
\newcommand{\qDold}{D_{3,q}^{0,\oldant}}
\newcommand{\qEold}{E_{3,q}^{0,\oldant}}
\newcommand{\gDold}{D_{3,g}^{0,\oldant}}
\newcommand{\gFold}{F_{3,g}^{0,\oldant}}
\newcommand{\gGold}{G_{3,g}^{0,\oldant}}
\newcommand{\gtoqAold}{A_{3,g \to q}^{0,\oldant}}
\newcommand{\gtoqDold}{D_{3,g \to q}^{0,\oldant}}
\newcommand{\QtogEold}{E_{3,\qp \to g}^{0,\oldant}}
\newcommand{\QtogGold}{G_{3,\qp \to g}^{0,\oldant}}

\newcommand{\calqAold}{{\cal A}_{3,q}^{0,\oldant}}
\newcommand{\calqDold}{{\cal D}_{3,q}^{0,\oldant}}
\newcommand{\calqEold}{{\cal E}_{3,q}^{0,\oldant}}

\newcommand{\calgFold}{{\cal F}_{3,g}^{0,\oldant}}
\newcommand{\calgGold}{{\cal G}_{3,g}^{0,\oldant}}
\newcommand{\calgtoqAold}{{\cal A}_{3,g \to q}^{0,\oldant}}
\newcommand{\calgtoqDold}{{\cal D}_{3,g \to q}^{0,\oldant}}
\newcommand{\calQtogEold}{{\cal E}_{3,\qp \to g}^{0,\oldant}}
\newcommand{\calQtogGold}{{\cal G}_{3,\qp \to g}^{0,\oldant}}

\newcommand{\qqA}{A_{3,q\bar{q}}^{0}}
\newcommand{\qgD}{D_{3,qg}^{0}}
\newcommand{\ggF}{F_{3,gg}^{0}}

\newcommand{\qgA}{A_{3,qg \to q\bar{q}}^{0}}
\newcommand{\ggD}{D_{3,gg \to gq}^{0}}
\newcommand{\qQE}{E_{3,q\qp \to qg}^{0}}
\newcommand{\gQG}{G_{3,g\qp \to gg}^{0}}

\newcommand{\qqAold}{A_{3,q\bar{q}}^{0,\oldant}}
\newcommand{\qgDold}{D_{3,qg}^{0,\oldant}}
\newcommand{\ggFold}{F_{3,gg}^{0,\oldant}}

\newcommand{\qgAold}{A_{3,qg \to q\bar{q}}^{0,\oldant}}
\newcommand{\ggDold}{D_{3,gg \to g\bar{q}}^{0,\oldant}}
\newcommand{\qQEold}{E_{3,q\qp \to qg}^{0,\oldant}}
\newcommand{\gQGold}{G_{3,g\qp \to gg}^{0,\oldant}}

\newcommand{\calqqA}{{\cal A}_{3,q\bar{q}}^{0}}
\newcommand{\calqgD}{{\cal D}_{3,qg}^{0}}
\newcommand{\calggF}{{\cal F}_{3,gg}^{0}}

\newcommand{\calqgA}{{\cal A}_{3,qg \to q\bar{q}}^{0}}
\newcommand{\calggD}{{\cal D}_{3,gg \to g\bar{q}}^{0}}
\newcommand{\calqQE}{{\cal E}_{3,q\qp \to qg}^{0}}
\newcommand{\calgQG}{{\cal G}_{3,g\qp \to gg}^{0}}

\title{\boldmath 
Initial-Final and Initial-Initial antenna functions for real radiation at next-to-leading order
}

\author[1]{Elliot Fox,}
\author[1,2]{Nigel Glover}

\affiliation[1]{Institute for Particle Physics Phenomenology,\\Department of Physics, \\Durham University, Durham, DH1 3LE, UK}
\affiliation[2]{Physik-Institut, Universität Zurich, Winterthurerstrasse 190, CH-8057 Zürich, Switzerland}

\emailAdd{elliot.fox@durham.ac.uk}
\emailAdd{e.w.n.glover@durham.ac.uk}

\preprint{IPPP/23/44, ZU-TH 47/23}

\abstract{The antenna subtraction method has achieved remarkable success in various processes relevant to the Large Hadron Collider. In Reference~\cite{paper2}, an algorithm was proposed for constructing real-radiation antenna functions for electron-positron annihilation, directly from specified unresolved limits, accommodating any number of real emissions. Here, we extend this algorithm to build antennae involving partons in the initial state, specifically the initial-final and initial-initial antennae. Using this extended algorithm, we explicitly construct all NLO QCD antenna functions and compare them with previously extracted antenna functions derived from matrix elements. Additionally, we rigorously match the integration of the antenna functions over the initial-final and initial-initial unresolved phase space with the previous approach, providing an independent validation of our results. The improved antenna functions are more compact and reduced in number, making them more readily applicable for higher-order calculations. 
}

\begin{document}
\maketitle
\flushbottom

\section{Introduction}

The Large Hadron Collider (LHC) presents an unprecedented opportunity to study a wide range of observables involving Higgs bosons, electroweak bosons, top quarks, and hadronic jets with remarkable accuracy. Through precise experimental measurements, we can directly probe the fundamental interactions of elementary particles at short distances, pushing the boundaries of our knowledge and gaining valuable insights into the fundamental forces governing the universe.

The exploration of LHC physics holds immense significance, especially in the absence of new particle discoveries. Scrutinizing LHC data with high precision allows us to detect even the slightest deviations from the predictions of the Standard Model (SM), which can profoundly impact our understanding of the natural world. Such small deviations have the potential to revolutionize our knowledge and steer us towards physics beyond the Standard Model, making precision phenomenology a critical aspect of the quest for new physics.

With the expected dataset from the High-Luminosity LHC, statistical uncertainties for many observables are likely to become negligible, enabling experimental accuracy at the percent level. Achieving similar percent-level accuracy for theoretical predictions necessitates further developments in fixed-order calculations, parton distribution functions, parton showers, and non-perturbative effect modeling. Ongoing progress in all of these areas will be crucial to meet the experimental precision that will be achieved at the LHC.

In recent years there has been enormous progress in fixed-order calculations, with many processes known to next-to-next-to-leading order (NNLO) in the strong-coupling expansion, and a few at next-to-next-to-next-to-leading order (N3LO).  Achieving such higher-order calculations demands careful attention due to the intricate interplay between real and virtual corrections across phase spaces with different multiplicity final states~\cite{Kinoshita,LeeNauenberg}. Implicit infrared divergences emerge due to unresolved real emissions, such as soft or collinear radiation. These divergences can only be canceled out by explicit poles arising from virtual graphs, achieved through integration over the relevant unresolved phase space. Currently, infrared subtraction schemes are considered one of the most elegant solutions to handle these subtleties, and ensure consistent and precise results.

\NG{Next-to-leading order (NLO) subtraction schemes such as Catani-Seymour dipole subtraction~\cite{Catani:1996vz,Catani:1996jh,Catani:2002hc} and FKS subtraction~\cite{Frixione:1995ms,Frixione:1997np} appeared in the mid-1990’s and have been fully developed for general collider processes. Both have been automated~\cite{Frederix:2008hu,Frederix:2009yq,Frederix:2010cj} and combined with automated one-loop matrix-element generators~\cite{madgraph:2011uj,Cascioli:2011va} to produce efficient parton-level event generators for fully-differential high-multiplicity processes. NLO matching schemes such as MC@NLO~\cite{Frixione:2002ik} and POWHEG~\cite{Nason:2004rx,Frixione:2007vw} have also been developed which combine the NLO fixed-order calculations with all-order parton-shower resummation to produce state-of-the-art multi-purpose event generators~\cite{powheg:2010xd,madgraph:2011uj,Bellm:2019zci,Sherpa:2019gpd,Bierlich:2022pfr}. Other methods have been established, notably the Nagy-Soper scheme~\cite{Nagy:2003qn,Bevilacqua:2013iha}, and others continue to be developed~\cite{Prisco:2020kyb,Bertolotti:2022ohq,Giachino:2023loc}. However, in the main, NLO subtraction is considered to be a solved problem.
}

\NG{At NNLO, the pattern of cancellation of infrared divergences across the different-multiplicity final states is much more complicated.  Following on from pioneering work by Anastasiou, Petriello and Melnikov~\cite{Anastasiou:2003gr} and Frixione~\cite{Frixione:2004is}, several subtraction schemes have been devised to compute the NNLO corrections to fully differential exclusive cross sections. Notable schemes
include Antenna subtraction~\cite{Gehrmann-DeRidder:2005btv, Gehrmann-DeRidder:2007foh, Currie:2013vh},
$q_T$-subtraction~\cite{Catani:2007vq}, Sector-Improved Residue subtraction~\cite{Czakon:2010td, Czakon:2014oma}, Nested soft-collinear subtraction~\cite{Caola:2017dug},
ColorFullNNLO subtraction~\cite{Somogyi:2006da, Somogyi:2006db, DelDuca:2016ily} and
Local Analytic subtraction~\cite{Magnea:2018hab}. Other subtraction schemes continue to be developed~\cite{Herzog:2018ily,TorresBobadilla:2020ekr}, while some ideas such as Projection-to-Born (P2B)~\cite{Cacciari:2015jma} sidestep the need for a general subtraction scheme.  P2B computes the NNLO corrections to fully differential exclusive cross sections related to a final state $X$ by combining the NLO calculation for differential cross sections for $X + j$ final states with the NNLO
corrections to the inclusive cross section for final state $X$. Because of the large increase in complexity compared to NLO, the implementation of these methods is currently done one process at a time.  They do not straightforwardly scale to higher multiplicities.  Work is therefore being done to extend or improve or automate these schemes, see for example~\cite{paper2, paper3, Gehrmann:2023dxm, Devoto:2023rpv}.
 }

\NG{At \NthreeLO, inclusive cross sections~\cite{Anastasiou:2015vya,Anastasiou:2016cez,Mistlberger:2018etf,Dreyer:2016oyx,Duhr:2019kwi,Duhr:2020kzd,Chen:2019lzz,Currie:2018fgr,Dreyer:2018qbw,Duhr:2020sdp,Duhr:2020seh}, as well as more differential calculations, have started to emerge~\cite{Dulat:2017prg,Dulat:2018bfe,Cieri:2018oms,Chen:2021isd,Chen:2021vtu,Billis:2021ecs,Chen:2022cgv,Neumann:2022lft,Camarda:2021ict,Chen:2022lwc,Baglio:2022wzu}, the latter mainly for $2 \to 1$ processes via the use of the Projection-to-Born method~\cite{Cacciari:2015jma} or $q_\mathrm{T}$-slicing techniques~\cite{Catani:2007vq} to promote established NNLO calculations to \NthreeLO.
The pattern of cancellation of infrared divergences across the different-multiplicity final states will be even more complicated than at NNLO making the construction of a more general subtraction scheme challenging.  The relevant infrared limits have been studied; the triple unresolved limits of tree amplitudes~\cite{Catani:2019nqv,DelDuca:1999iql,DelDuca:2019ggv,DelDuca:2020vst,DelDuca:2022noh}, the double unresolved limits of one-loop amplitudes~\cite{Catani:2003vu,Sborlini:2014mpa,Badger:2015cxa,Zhu:2020ftr,Catani:2021kcy,Czakon:2022fqi}, and the single unresolved limits of two-loop amplitudes~\cite{Bern:2004cz,Badger:2004uk,Duhr:2014nda,Li:2013lsa,Duhr:2013msa} have all been studied and could in principle lead to an N3LO subtraction scheme.
We note that the first steps towards an \NthreeLO antenna subtraction scheme have been taken in Refs.~\cite{Jakubcik:2022zdi,Chen:2023fba,Chen:2023egx}. Nevertheless, at the moment, calculations for higher multiplicities are currently hindered by the lack of process-independent \NthreeLO subtraction schemes.}

The Antenna subtraction scheme is a highly successful method for fully-differential NNLO calculations in perturbative QCD. It was initially proposed for massless partons in electron-positron annihilation at NNLO~\cite{Gehrmann-DeRidder:2005btv,Gehrmann-DeRidder:2005alt,Gehrmann-DeRidder:2005svg,Gehrmann-DeRidder:2007foh} and then extended to treat initial-state radiation with initial-state hadrons at NLO~\cite{Daleo:2006xa} and later at NNLO~\cite{Daleo:2009yj,Pires:2010jv,Boughezal:2010mc,Gehrmann:2011wi,Gehrmann-DeRidder:2012too,Currie:2013vh}. Moreover, the framework has been applied to heavy particle production~\cite{Gehrmann-DeRidder:2009lyc,Abelof:2011ap,Bernreuther:2011jt,Abelof:2011jv,Abelof:2012bga,Abelof:2012rv,Bernreuther:2013uma,Dekkers:2014hna} and adapted to antenna-shower algorithms~\cite{Gustafson:1987rq,Lonnblad:1992tz,Giele:2007di,Giele:2011cb,Fischer:2016vfv,Brooks:2020upa} enabling higher-order corrections~\cite{Li:2016yez} and the first approaches to fully-differential NNLO matching~\cite{Campbell:2021svd}. \NG{It is based around tree antennae for single ($X_3^0$) and double ($X_4^0$) radiation, and one-loop antennae for single radiation ($X_3^1$).  Combinations of these three types of antennae, together with a wide-angle soft term, are sufficient to build subtraction terms at NNLO.  A particular feature is that the single unresolved $X_3^0$ antenna are heavily reused in constructing NNLO subtraction terms. Simplifying and streamlining the $X_3^0$ antenna will have a knock on effect at NNLO and beyond.} 

In its original formulation, the antennae were constructed from simple matrix elements which have the desired infrared singularities. By construction, the factorisation properties of matrix elements guaranteed that the subtraction term would match the infrared limit of the full matrix element. The direct extraction of antenna functions from matrix elements elegantly bypassed the complexities of combining subtraction terms for various multiple-soft and/or collinear limits. However, this approach gave rise to two issues.

Firstly, when dealing with double-real radiation antenna functions derived from matrix elements, it was challenging to unequivocally identify the hard radiators, especially for partonic configurations involving gluons. To address this, sub-antenna functions were introduced. Unfortunately, constructing sub-antenna functions at NNLO proved to be extremely cumbersome, often leading to unphysical denominators that hindered analytic integration. In many cases, only the analytic integrals for the full antenna functions were available. Consequently, assembling antenna-subtraction terms necessitated a process where sub-antenna functions were combined to reconstruct the full antenna functions before integration.

Secondly, NNLO antenna functions could contain spurious limits, necessitating the use of explicit counter terms to remove them. This introduced further spurious singularities, creating a chain of cross-dependent subtraction terms that lacked any connection to the actual singularity structure of the underlying process.

To overcome these challenges and streamline the antenna-subtraction scheme, Refs.~\cite{paper2,paper3} introduced a set of design principles to algorithmically construct antenna functions for final-state particles directly from the relevant infrared limits, while properly accounting for the overlap between different limits.
The algorithm applied to antenna with final-state particles ensures that
\begin{enumerate}[I.]
\item each antenna function has exactly two hard particles (``radiators'') which cannot become unresolved;
\item each antenna function captures all (multi-)soft limits of its unresolved particles;
\item where appropriate, (multi-)collinear and mixed soft and collinear limits are decomposed over ``neighbouring'' antennae;
\item antenna functions do not contain any spurious (unphysical) limits;
\item antenna functions only contain singular factors corresponding to physical propagators; and
\item where appropriate, antenna functions obey physical symmetry relations (such as line reversal).
\end{enumerate}

The iterative algorithm for constructing real radiation antenna functions \NG{for massless partons} with a particular set of 
soft and/or collinear singular limits pertaining to the two hard radiators is described in detail in Refs.~\cite{paper2,paper3}.  It relies on:
\begin{itemize}
    \item a list of “target functions”, which in the following we will denote by $L_i$, which capture the behaviour of the colour-ordered matrix element squared in the given unresolved limit and are taken as input to the algorithm,
    \item a set of “down-projectors” $\PPdown_i$ which maps the invariants of the full phase space into the subspace relevant for limit $L_i$,
    \item a set of “up-projectors” $\PPup_i$ which restores the full antenna phase space. Note that the down-projectors $\PPdown_i$ and up-projectors $\PPup_i$ are typically not inverse to each other, as down-projectors destroy information about less-singular and finite pieces.
\end{itemize}

In this paper we want to generalise this algorithm to include the cases when one or both of the hard radiators are in the initial state.  These are the initial-final and initial-initial antenna respectively. We focus on the single unresolved (NLO) antenna functions \NG{for massless partons}.   

The paper is structured as follows.
In Section~\ref{sec:NLOantennae} we classify the various types of initial-final and initial-initial antennae needed for NLO calculations.  
We introduce the appropriate ingredients for the algorithm in \ref{sec:building}, defining the relevant limits and their associated $\PPdown$ and $\PPup$ operators.  Sections~\ref{sec:IF} and \ref{sec:II} give explicit forms for the initial-final and initial-initial antennae, as well as their analytic integration over the appropriate phase space.  At each stage, we compare with the antennae derived from matrix elements~\cite{Daleo:2006xa}. Finally, we conclude and give an outlook on further work in Section~\ref{sec:outlook}.

\section{NLO antennae with particles in the initial state}
\label{sec:NLOantennae}

In the antenna subtraction scheme, antenna functions are used to subtract specific sets of unresolved singularities, so that a typical subtraction term has the form
\begin{equation} \label{eqn:subterm}
    X_{n+2}^\ell(i_1^h,i_3,\ldots,i_{n+2},i_2^h) 
    \modsq{\calM(\ldots,I_1^h,I_2^h,\ldots)} \, ,
\end{equation}
where $X_{n+2}^\ell$ represents an $\ell$-loop, $(n+2)$-particle antenna,
$i_1^h$ and $i_2^h$ represent the hard radiators, and $i_3$ to $i_{n+2}$ denote the $n$ unresolved particles.   
As the hard radiators may either be in the initial or in the final-state, final-final (FF), initial-final (IF), and initial-initial (II) configurations need to be considered in general. $\calM$ is the reduced matrix element, with $n$ fewer particles and where $I_1^h$ and $I_2^h$ represent the particles obtained through an appropriate mapping,
\begin{align}
\{ p_{i_1},p_{i_3},\ldots,p_{i_2} \} \mapsto \{ p_{I_1}, p_{I_2} \}  
\end{align}
with $p_i^{\mu}$ representing the four-momentum of particle $i$. At NLO, antennae have $n=1$ and $\ell=0$, at NNLO one needs antennae with $n=2,~\ell = 0$ and with $n=1,~\ell=1$, while at \NthreeLO, one needs antennae with $n=3,~\ell = 0$, with $n=2,~\ell = 1$ and with $n=1,~\ell = 2$. 
In the original formulation of the antenna scheme, the antennae were constructed from simple matrix elements which have the desired singularities. By construction, the factorisation properties of matrix elements guaranteed that the subtraction term would match the infrared limit of the full matrix element. 
\NG{As mentioned in the Introduction, we focus on the tree-level single unresolved antenna for massless partons.  The extension to massive partons is straightforward.}

The tree-level three particle antenna for particles of type $a$, $b$ and $c$ in the final-state is denoted by $X_3^0(a^h,b,c^h)$.  The antenna describes the infrared singularities when particle $b$ is unresolved: the soft $b$ singularity (if it exists), as well as (parts of) the collinear singularities $a||b$ and $b||c$.  

The two hard ``radiators'', $a^h$ and $c^h$, cannot become unresolved. Therefore, the antenna should not contain singularities when $a$ or $c$ are soft or when $a||c$.  

To see how this happens, let us focus on the soft $c$ singularity.  The argument for soft $a$ is similar.  

The soft $c$ singularity is avoided by including only the parts of the $bc$ splitting function that are not singular when $c$ is soft.  We denote this as $b||c^h$, and divide up the collinear splitting function across two antennae, $X_3^0(a^h,b,c^h)$ and $X_3^0(b^h,c,d^h)$, such that the full collinear limit is recovered. 
In this case, the full $b||c$ collinear limit would be obtained by two subtraction terms,
\begin{equation} \label{eqn:subterm2}
    \X(a^h,b,c^h) 
    \modsq{\calM(\ldots,A^h,C^h,\ldots)} \, 
    +\X(b^h,c,d^h) 
    \modsq{\calM(\ldots,B^h,D^h,\ldots)} \, .
\end{equation}
Here $\{A^h, C^h\}$ and $\{B^h,D^h\}$ are obtained by applying the antenna mapping to $\{a^h,b,c^h\}$ and $\{b^h,c,d^h\}$ respectively.

We note that when particles are written in order of colour connection, $X_3^0(a^h,b,c^h)$ will contain the required soft singularities $\sim 1/s_{ab}s_{bc}$, as well as the collinear singularities $1/s_{ab}$ and $1/s_{bc}$, while avoiding the $1/s_{ac}$ collinear singularity.  

The full list of final-final antennae is given in Table~\ref{tab:X30FF} in the form $\X(i_{a}^h, j_{b}, k_{c}^{h})$ where $i$, $j$ and $k$ represent the momenta of particles $a$, $b$ and $c$ respectively.  There are five distinct antenna configurations corresponding to the colour-connected strings for the $qg\bar{q}$, $qgg$, $q\qpbar\qp$, $ggg$ and $g\qpbar\qp$ particle assignments. Noting that under charge conjugation and colour line-reversal
\begin{equation}
\label{eq:chargeconjugation}
    \X(k_{\bar{c}}^h, j_{\bar{b}}, i_{\bar{a}}^h) = \X(i_a^h,j_b,k_c^h), 
\end{equation}
these antenna also describe $gg\bar{q}$, $\qpbar\qp\bar{q}$ and $\qpbar\qp g$ colour-connected strings.  
Expressions for each of these antennae together with the integrals over the final-final antenna phase space are given in Ref.~\cite{paper2}.
Frequently, we drop explicit reference to the particle labels in favour of a specific choice of $X$ according to Table~\ref{tab:X30FF}. 

\begin{table}[ht]
\centering
\begin{tabular}{lclcl}
\bf{Final-Final Antennae}  & & &&\\
\\
$A_3^0(i^h,j,k^h)$  & $\equiv $ &$ \X(i_q^h,j_g,k_{\bar{q}}^h)$      & &\\
$D_3^0(i^h,j,k^h)$  & $\equiv $ &$ \X(i_q^h,j_g,k_g^h)$              & $\equiv $ &$ \X(k_g^h,j_g,i_{\bar{q}}^h)$\\
$E_3^0(i^h,j,k^h)$  & $\equiv $ &$ \X(i_q^h,j_{\qpbar},k_{\qp}^h)$   & $\equiv $ &$ \X(k_{\qpbar}^h,j_{\qp},i_{\bar{q}}^h)$\\ 
$F_3^0(i^h,j,k^h)$  & $\equiv $ &$ \X(i_g^h,j_g,k_g^h)$              & &\\
$G_3^0(i^h,j,k^h)$  & $\equiv $ &$ \X(i_g^h,j_{\qpbar},k_{\qp}^h)$   & $\equiv $ &$ \X(k_{\qpbar}^h,j_{\qp},i_{g}^h)$\\\\
\end{tabular}
\caption{Identification of Final-Final $X_3^0$ antenna according to the particle type. These antennae only contain singular limits when the particle with momentum $j$ is unresolved. The antenna configurations in the third column are related to the functions in the first column by charge-conjugation as in Eq.~\eqref{eq:chargeconjugation}. }
\label{tab:X30FF}
\end{table}

We obtain antennae with initial-state particles by crossing one (or two) of the particles into the initial state. We denote initial-state particles with a hat, noting that the crossing should also be applied to the final-state particle such that, for example, $\hat{i}_{Q}$ denotes an initial-state anti-quark. 

We define antennae of {\bf Type 1} by crossing one (or two) of the hard radiators into the initial state.  There are two initial-final configurations,
$$X_3^0(\widehat{a},b,c^h),\qquad 
X_3^0(a^h,b,\widehat{c}),$$   
and one initial-initial case,
$$X_3^0(\widehat{a},b,\widehat{c}).$$
Just as for final-final antennae, these antennae potentially have singularities when $b$ is soft, or when $b$ is collinear with either of the hard radiators.

We define antennae of {\bf Type 2} by crossing the unresolved particle into the initial state.  For initial-initial antenna, we also cross one of the hard radiators into the initial state.  We ensure that the number of hard radiators is two.  There are two configurations for the initial-final,
$$X_3^0(a^h,\widehat{b},c),\qquad 
X_3^0(a,\widehat{b},c^h),$$   
and two for the initial-initial cases,
$$X_3^0(\widehat{a},\widehat{b},c),\qquad
X_3^0(a,\widehat{b},\widehat{c})$$
respectively. These antennae do not have any soft singularities. They have no singularities when $\widehat{b}$ is collinear to the other hard radiator ($\widehat{a}$ or $\widehat{c}$), but do potentially have singularities when $\widehat{b}$ is collinear with the unresolved particle ($c$ or $a$).

One needs to be careful about how the antenna mapping affects the identity of the initial-state particles.  We therefore discriminate between two cases:
\begin{enumerate}
    \item The particle type resulting from clustering an initial-state and final-state particle is {\bf the
    same as} the particle type in the initial state, i.e. $\hat{a}b \to \hat{a}$.
    For example, $\hat{q} g \to \hat{q}$.
    We label this type of antenna as {\bf Identity Preserving (IP)}.
    \item The particle type resulting from clustering an initial-state and final-state particle is {\bf different from} the particle type in the initial state, i.e. $\hat{a}b \to \hat{c}$.
    For example, $\hat{g} q \to \hat{q}$ or $\hat{q}\bar{q}\to g$.
    We label this type of antenna as {\bf Identity Changing (IC)}. Each antenna should include at most one identity change. 
\end{enumerate}

We further arrange that Type 1 antennae are IP, while Type 2 antennae are IC.
As an example, consider the subtraction term for a particular colour-ordered matrix element with particle $\hat{c}$ in the initial state, $\modsq{\calM(\ldots,a,b,\widehat{c},d,e,\ldots)}$.   Focussing on the colour-string triplets, $ab\hat{c}$, $b\hat{c}d$, $\hat{c}de$ that contain collinear singularities associated with either $b||\hat{c}$ or $\hat{c}||d$, there are four possible contributions. 
\begin{enumerate}
    \item $\textrm{soft}~b, \IP(b||\widehat{c})$ $$\X(a^h,b,\widehat{c}) 
    \modsq{\calM(\ldots,A^h,\widehat{C}^h,d,e,\ldots)}$$

    This term would contribute if the particle type of $b\widehat{c}$ {\bf is the same} as $\widehat{c}$.  It would describe the soft $b$ singularity and the IP  $b||\widehat{c}$ collinear singularity. In this case, we would include the full $b||\widehat{c}$ splitting function (since initial particle $c$ cannot be soft).  Note that this contribution also contains the $a^h||b$ singularity.

    \item $\IC(b||\widehat{c})$
    $$\X(b,\widehat{c},d^h)
    \modsq{\calM(\ldots,a,\widehat{B}^h,D^h,e,\ldots)}$$

    This term would contribute if the particle type of $b\widehat{c}$ ($\widehat{B}$)  {\bf is NOT the same} as $\widehat{c}$.  It would describe IC  $b||\widehat{c}$ collinear singularity.  This term contains {\bf NO} $\widehat{c}||d^h$ singularity.

    \item $\IC(\widehat{c}||d)$
    $$\X(b^h,\widehat{c},d)  
    \modsq{\calM(\ldots,a,B^h,\widehat{D}^h,e,\ldots)}$$

    This term would contribute if the particle type of $\widehat{c}d$ ($\widehat{D}$)  {\bf is NOT the same} as $\widehat{c}$.  It would describe IC  $\widehat{c}||d$ collinear singularity.  This term contains {\bf NO} $b^h||\widehat{c}$ singularity.

    \item $\textrm{soft}~d, \IP(\widehat{c}||d)$
    $$\X(\widehat{c},d,e^h) 
    \modsq{\calM(\ldots,a,b,\widehat{C}^h,E^h,\ldots)}$$

    This term would contribute if the particle type of $\widehat{c}d$ ($\widehat{C}$) {\bf is the same} as $\widehat{c}$.  It would describe the soft $d$ singularity and the IP  $\widehat{c}||d$ collinear singularity. In this case, we would include the full $\widehat{c}||d$ splitting function (since initial particle $c$ cannot be soft). Note that this contribution also contains the $d||e^h$ singularity.

\end{enumerate}

\begin{table}[ht]
\centering
\begin{tabular}{lclcl}
\bf{Initial-Final Antennae}  & & &&\\
\\
\bf{Identity preserving}  & & & &\\
$\qA(\widehat{i},j,k^h)$ 
& $\equiv $ & $\X(\widehat{i}_{q},j_g,k_{\bar{q}}^h)$ 
& $\equiv $ & $\X(k_{q}^h,j_g,\widehat{i}_{\bar{q}})$
\\
$\qD(\widehat{i},j,k^h)$ 
& $\equiv $ & $\X(\widehat{i}_{q},j_g,k_{g}^h)$       
& $\equiv $ & $\X(k_{g}^h,j_g,\widehat{i}_{\bar{q}})$  
\\
$\qE(\widehat{i},j,k^h)$
& $\equiv $ &  $\X(\widehat{i}_{q},j_{\qpbar},k_{\qp}^h)$  
& $\equiv $ &  $\X(k_{\qpbar}^h,j_{\qp},\widehat{i}_{\bar{q}})$  
\\
$\gD(k^h,j,\widehat{i})$ 
& $\equiv $ &  $\X(k_{q}^h,j_{g},\widehat{i}_{g})$ 
& $\equiv $ &  $\X(\widehat{i}_{g},j_{g},k_{\bar{q}}^h)$ 
\\
$\gF(\widehat{i},j,k^h)$ 
& $\equiv $ &  $\X(\widehat{i}_{g},j_{g},k_{g}^h)$ 
& $\equiv $ &  $\X(k_{g}^h, j_{g}, \widehat{i}_{g})$ 
\\
$\gG(\widehat{i},j,k^h)$ 
& $\equiv $ &  $\X(\widehat{i}_{g},j_{\qpbar},k_{\qp}^h)$ 
& $\equiv $ &  $\X(k_{\qpbar}^h,j_{\qp},\widehat{i}_{g})$ 
\\
\bf{Identity changing} & \\
$\gtoqA(j,\widehat{i},k^h)$
& $\equiv $ &  $\X(j_{q},\widehat{i}_{g},k_{\bar{q}}^h)$ 
& $\equiv $ &  $\X(k_{q}^h,\widehat{i}_{g},j_{\bar{q}})$ 
\\
$\gtoqD(j,\widehat{i},k^h)$
& $\equiv $ &  $\X(j_{q},\widehat{i}_{g},k_{g}^h)$ 
& $\equiv $ &  $\X(k_{g}^h,\widehat{i}_{g},j_{\bar{q}})$ 
\\
$\QtogE(k^h,\widehat{i},j)$
& $\equiv $ &  $\X(k_{q}^h,\widehat{i}_{\qpbar},j_{\qp})$ 
& $\equiv $ &  $\X(j_{\qpbar},\widehat{i}_{\qp},k_{\bar{q}}^h)$ 
\\
$\QtogG(k^h,\widehat{i},j)$
& $\equiv $ &  $\X(k_{g}^h,\widehat{i}_{\qpbar},j_{\qp})$ 
& $\equiv $ &  $\X(j_{\qpbar},\widehat{i}_{\qp},k_{g}^h)$  
\\
\end{tabular}
\caption{Identification of the Initial-Final $X_3^0$ antenna according to the particle type. These antennae only contain singular limits when the particle with momentum $j$ is unresolved. The initial-state particle carries momentum $i$. The particle assignments in functions in the first column are by convention equal to those in the second column.  The antenna configurations in the third column are related to the functions in the first column by charge-conjugation as in Eq.~\eqref{eq:chargeconjugation}.}
\label{tab:X30IF}
\end{table}

The identification of the Initial-Final antennae according to particle type is listed in Table~\ref{tab:X30IF}.  
As for the Final-Final antennae, we drop explicit reference to the particle labels in favour of a specific choice of $X$ according to Table~\ref{tab:X30IF}.
We systematically use the momentum set $\{\widehat{i}, j, k^h\}$ with the initial momentum $\widehat{i}$, hard radiator with momentum $k^h$, and the unresolved particle with momentum $j$.  This means that soft singularities are characterised by $1/s_{ij}s_{jk}$ and collinear singularities by $1/s_{ij}$ and $1/s_{jk}$ in the IP antenna, while the IC antenna only have singularities proportional to $1/s_{ij}$.
Since the initial-state particles cannot be soft, we will also drop the superscript $h$ on the initial-state (hatted) momenta.

\begin{table}[ht]
\centering
\begin{tabular}{lclcl}
\bf{Initial-Initial Antennae}  & & &&\\
\\
\bf{Identity preserving} & \\
$\qqA(\widehat{i},j,\widehat{k})$ 
& $\equiv $ &  $\X(\widehat{i}_{q},j_{g},\widehat{k}_{\bar{q}})$ 
& &   
\\
$\qgD(\widehat{i},j,\widehat{k})$
& $\equiv $ &  $\X(\widehat{i}_{q},j_{g},\widehat{k}_{g})$  
& $\equiv $ &  $\X(\widehat{k}_{g}, j_{g}, \widehat{i}_{\bar{q}})$  
\\
$\ggF(\widehat{i},j,\widehat{k})$ 
& $\equiv $ &  $\X(\widehat{i}_{g},j_{g},\widehat{k}_{g})$ 
& &  
\\
\bf{Identity changing} & \\
$\qgA(\widehat{i},\widehat{k},j)$
& $\equiv $ &  $\X(\widehat{i}_{q},\widehat{k}_{g},j_{\bar{q}})$ 
& $\equiv $ &  $\X(j_q,\widehat{k}_{g},\widehat{i}_{\bar{q}})$ 
\\
$\ggD(\widehat{i},\widehat{k},j)$
& $\equiv $ &  $\X(j_q,\widehat{k}_{g},\widehat{i}_{g})$ 
& $\equiv $ &  $\X(\widehat{i}_{g},\widehat{k}_{g},j_{\bar{q}})$ 
\\
$\qQE(\widehat{i},\widehat{k},j)$ 
& $\equiv $ &  $\X(\widehat{i}_{q},\widehat{k}_{\qpbar},j_{\qp})$ 
& $\equiv $ &  $\X(j_{\qpbar},\widehat{k}_{\qp},\widehat{i}_{\bar{q}})$ 
\\
$\gQG(\widehat{i},\widehat{k},j)$
& $\equiv $ &  $\X(\widehat{i}_{g},\widehat{k}_{\qpbar},j_{\qp})$ 
& $\equiv $ &  $\X(j_{g},\widehat{k}_{\qp},\widehat{i}_{\bar{q}})$ 
\\
\end{tabular}

\caption{Identification of the Initial-Initial $X_3^0$ antenna according to the particle type. These antennae only contain singular limits when the particle with momentum $j$ is unresolved. The initial-state particles carry momenta $i$ and $k$. The antenna configurations in the third column are related to the functions in the first column by charge-conjugation as in Eq.~\eqref{eq:chargeconjugation}.}
\label{tab:X30II}
\end{table}

The identification of the Initial-Initial antennae according to particle type is listed in Table~\ref{tab:X30II}.  
As usual, we drop explicit reference to the particle labels in favour of a specific choice of $X$ according to Table~\ref{tab:X30II}.
We systematically use the momentum set $\{\widehat{i}, j, \widehat{k}\}$ with the initial momenta $\widehat{i}$ and $\widehat{k}$ and the unresolved particle carries momentum $j$.  This means that soft singularities are characterised by $1/s_{ij}s_{jk}$ and collinear singularities by $1/s_{ij}$ and $1/s_{jk}$ in the IP antenna, while the IC antenna only have singularities proportional to $1/s_{jk}$.
 As for the IF antennae, since the initial-state particles cannot be soft, we will also drop the superscript $h$ on the initial-state (hatted) momenta.  

\section{Building the antennae}
\label{sec:building}

A core part of our algorithm is the definition of down-projectors into singular limits with corresponding up-projectors into the full phase space. In each step of the construction, down-projectors are needed to identify the overlap of the so-far constructed antenna function with the target function of the respective unresolved limit, whereas up-projectors are required to re-express the subtracted target function in terms of antenna invariants. In this way, the full (accumulated) antenna function can be expressed solely in terms of $n$-particle invariants, and is therefore valid in the full phase space. By choosing the up-projectors judiciously, the antenna function can furthermore be expressed exclusively in terms of physical propagators. As alluded to above, down-projectors $\PPdown$ and up-projectors $\PPup$ are not required to be inverse to each other.

The primary objective of the algorithm is to construct antenna functions for (multiple-)real radiation that encompass singular limits associated with precisely two hard radiators, along with an arbitrary number of additional particles allowed to remain unresolved. To achieve this goal, each limit is specified by a "target function", denoted as $L_i$. These target functions are crucial in capturing the behavior of the color-ordered matrix element squared under the specified unresolved limit, and serve as input for the algorithm. For the three-particle antennae constructed in this paper, the relevant limits are:
\begin{itemize}
    \item soft final-state particle,
    \item two final-state particles are collinear,
    \item a final-state particle is collinear with an initial-state particle.
\end{itemize}
It is the third case that is new in this paper. 

Although the target functions might encompass process-dependent azimuthal terms, for the sake of simplicity, we will focus solely on azimuthally-averaged functions as in Ref.~\cite{Daleo:2006xa}. This is not necessarily a limitation, and one could in principle include the azimuthal terms if desired.

Note that we use the Lorentz-invariant definition of the invariants,
\begin{equation}
s_{i\dots n} = (p_i+\ldots+p_n)^2.
\end{equation}
For final-final configurations, all invariants are positive.
However, under crossing, $p_i \to -p_i$ for IF and $\{p_i \to -p_i, p_k \to - p_k\}$ for II, so some of the invariants become negative:
\begin{eqnarray}
    \textrm{IF}: && \sij < 0, \qquad \sjk > 0, \qquad \sik < 0, \qquad \sijk < 0,\\
    \textrm{II}: && \sij < 0, \qquad \sjk < 0, \qquad \sik > 0, \qquad \sijk > 0.
\end{eqnarray}

\subsection{The soft projectors}

The soft down-projector given in Ref.~\cite{paper2} is defined by the mapping
\begin{equation}
\PSdown_{j}:
\begin{cases}
\sij\rightarrow \lambda \sij\\
\sjk\rightarrow \lambda \sjk\\
\sijk \rightarrow \sik
\end{cases}
\end{equation}
and we keep only terms of order $\lambda^{-2}$. The corresponding $\PSup_{j}$ is just the trivial mapping which leaves all variables unchanged. The unresolved soft particle must always be in the final state.  However, the hard radiators $i$ and $k$ may be in either the final or the initial state.  Crossing from the final to the initial state does not affect the power counting.  Therefore this is the only soft projector we will need.

The only particle that produces a non-trivial limit is the gluon, where the relevant soft limit is the eikonal factor
\begin{equation}
\label{eq:softlimit}
S_{g}(i^{h},j_{g},k^{h})\equiv 
\frac{2\sik}{\sij\sjk}.
\end{equation}
Under crossing, the eikonal factor is unchanged,
\begin{equation}
    S_{g}(\widehat{i},j_{g},k^{h}) = S_{g}(\widehat{i},j_{g},\widehat{k}) = S_{g}(i^{h},j_{g},k^{h}).
\end{equation}

\subsection{The collinear projectors}
\subsubsection{FF: Both collinear particles in final state}

When the unresolved final-state particle $j$ becomes collinear to a final-state hard radiator particle $i$, the collinear projectors are given by~Ref.~\cite{paper2}
\begin{equation}
\label{eq:CijdownFF}
\PCdownFF_{ij}:
\begin{cases}
\sij\rightarrow \lambda \sij\\
\sik \rightarrow (1-x_{j})(\sik+\sjk)\\
\sjk \rightarrow x_{j}(\sik+\sjk)\\
\sijk \rightarrow (\sik+\sjk)
\end{cases}
\end{equation}
where we only keep terms of order $\lambda^{-1}$. The momentum fraction $x_{j}=\frac{\sjk}{\sik+\sjk}$ is defined with respect to the spectator third particle in the antenna $k$.  The corresponding up-projector is 
\begin{equation}
\label{eq:CijupFF}
\PCupFF_{ij}:
\begin{cases}
x_{j} \rightarrow \frac{\sjk}{\sijk}\\
(1-x_{j})\rightarrow \frac{\sik}{\sijk}\\
\sik+\sjk\rightarrow\sijk.
\end{cases}
\end{equation}

The projectors when $j$ becomes collinear to the final-state hard radiator particle $k$ are obtained by exchanging the roles of $i$ and $k$,
\begin{align}
 \PCupFF_{kj} &=  \PCupFF_{ij} \qquad i \leftrightarrow k,\\
 \PCdownFF_{kj} &=  \PCdownFF_{ij} \qquad i \leftrightarrow k.
\end{align}

In the final-final collinear case, the relevant limits are given by the splitting functions $P_{ab}(i^h,j)$, which are {\em not singular} in the limit where the hard radiator $a$ becomes soft, and are related to the usual spin-averaged splitting functions, cf.~\cite{Altarelli:1977zs,Dokshitzer:1977sg}, by, 
\begin{align} 
\label{eqn:Pqg}
\Pqg(i^h,j) &= \frac{1}{s_{ij}} \Pqg(\xj) \\
\Pqg(i,j^h) &= 0,\\
\label{eqn:Pqq}
\Pqq(i^h,j) &= \frac{1}{s_{ij}} \Pqq(\xj),\\
\Pqq(i,j^h) &= \frac{1}{s_{ij}} \Pqq\omxj,\\
\label{eqn:Pgg}
\Pgg(i^h,j) &= \frac{1}{s_{ij}} \PggS(\xj)\hfill\\
\Pgg(i,j^h) &= \frac{1}{s_{ij}}\PggS\omxj
\end{align}
with
\begin{eqnarray}
\Pqg(\xj) &=& \left(\frac{2\omxj}{\xj} + \ome \xj \right),\\
\Pqq(\xj) &=& \left( 1 -\frac{2\omxj\xj}{\ome} \right) = \Pqq\omxj,\\
\label{eq:PggS}
\PggS(\xj)&=& \left( \frac{2\omxj}{\xj} + \xj \omxj \right), 
\end{eqnarray}
and
\begin{equation}
\PggS(\xj) + \PggS\omxj \equiv \Pgg(\xj).
\end{equation}

\NG{
{\bf Azimuthal spin correlations:}\\
In general, the collinear limits of matrix elements will contain spin correlations, and hence are accurately reproduced using spin-dependent splitting functions.
For example, the spin dependent gluon splitting function is~\cite{Catani:1996vz},
\begin{eqnarray}
P_{gg}^{\mu\nu}(\xj)=2\left[-g^{\mu\nu}\left(\frac{\xj}{1-\xj}+\frac{1-\xj}{\xj}\right)-2(1-\epsilon)\xj(1-\xj)
\frac{k_{\perp}^\mu k_{\perp}^\nu}{k_{\perp}^2}\right],
\end{eqnarray}
where $k_{\perp}$ represents the momentum transverse to the collinear direction. 
The spin-averaged splitting functions that we use to build the antenna are obtained by integrating these spin-dependent splitting functions over the azimuthal angle of the plane containing the collinear particles with respect to the collinear direction. This means that in a point-by-point check, the subtraction terms based on the spin-averaged splitting functions will not correctly reproduce the 
azimuthal terms present in the matrix elements.}

\NG{
However, we note that the angular terms related to each other by a rotation of the system of unresolved partons precisely cancel~\cite{Weinzierl:2006wi,Gehrmann-DeRidder:2007foh}.  It can be shown that the angular terms are proportional to $\cos(2\phi+\alpha)$
where $\phi$ is the azimuthal angle around the collinear direction.  Therefore by combining two phase space points with azimuthal angles $\phi$ and $\phi+\pi/2$ and all other coordinates equal, the azimuthal correlations drop out.  When particles $i$ and $j$ are in the final state, this can be achieved by rotating particles $i$ and $j$ by $\pi/2$ about the collinear direction. 
In initial-final configurations produced when $p_i^\mu \to p^\mu+p_j^\mu$ for $i=1,2$ and with $i||j$ and $p^2=0$, the phase space points are again be related by azimuthal rotations of $\pi/2$.  This has the consequence of rotating $p_i^\mu$ off the beam axis and therefore has to be compensated by a Lorentz boost. Once this averaging takes place, the collinear limit will be accurately described by the spin-averaged splitting function. We verify this claim in Section~\ref{sec:validation}.}

\subsubsection{Mixed initial/final collinear limit.}

If we have a final-state parton with momentum $p_{j}$ becoming collinear with an initial-state hard radiator with momentum $p_{i}$, then
\begin{equation}
\begin{split}
&p_{j}\rightarrow \zj p_{i},\\
&p_{ij}:=(p_{i}-p_{j})\rightarrow \omzj p_{i},\\
&\sik\rightarrow \frac{\sijk}{\omzj},\\
&\sjk\rightarrow  -\frac{\zj\sijk}{\omzj}.
\end{split}
\end{equation}
In this case, the relevant limits are given by the splitting functions
\begin{align}
    P_{gq\leftarrow Q}(\widehat{i},j) &=\frac{-1}{\sij} P_{gq\leftarrow Q}(\zj),\\
    P_{qg\leftarrow Q}(\widehat{i},j) &=\frac{-1}{\sij} P_{qg\leftarrow Q}(\zj),\\
    P_{q\bar{q}\leftarrow G}(\widehat{i},j) &=\frac{-1}{\sij} P_{q\bar{q}\leftarrow G}(\zj),\\
    P_{gg\leftarrow G}(\widehat{i},j) &=\frac{-1}{\sij} P_{gg\leftarrow G}(\zj),
\end{align}
and
\begin{equation}
P_{ab \leftarrow c} (j, \widehat{i}) = P_{ba \leftarrow c} ( \widehat{i},j)
\end{equation}
with
\begin{align}
P_{gq\leftarrow Q}(\zj) 
&= \frac{1}{\omzj\ome}P_{qg}\omzj 
=\frac{2\zj}{\omzj^{2}\ome}+1,\\
P_{qg\leftarrow Q}(\zj) 
&= \frac{1}{\omzj}P_{qg}(\zj) 
=\frac{2}{\zj}+\frac{\ome \zj}{\omzj}, \\
\label{Pqqfull}
P_{q\bar{q}\leftarrow G}(\zj) 
&= \frac{\ome}{\omzj}P_{q\bar{q}}(\zj) 
=\frac{\ome}{\omzj}-2\zj, \\
\label{Pggfull}
P_{gg\leftarrow G}(\zj) 
&= \frac{1}{\omzj}P_{gg}(\zj) 
=\frac{2}{\zj}+\frac{2\zj}{\omzj^{2}}+2\zj.
\end{align}
Here $P_{ab\leftarrow C}$ is the splitting of the initial-state $C$ into parton $a$ entering the hard process and parton $b$ radiated.
We note that $\zj = -\sjk/\sik$.  When the spectator particle $k$ is in the initial state, then $\sik$ can never be small.   However, if $k$ is in the final state, then $\sik$ may be small, and we have to avoid introducing fake singularities in this limit.  This can happen when we encounter either of the $P_{q\bar{q}\leftarrow G}$ or $P_{gg\leftarrow G}$ splitting functions. 
Therefore, we generate mappings that discriminate between the two cases (IF) when $k$ is in the final state and (II) when $k$ is in the initial state.

\subsubsection*{(IF) Spectator in final state.}

The Initial-Final collinear down-projector when particles $\widehat{i}$ and $j$ are collinear and the spectator $k$ is in the final state is given by
\begin{equation}
\label{CijdownIF}
\PCdownIF_{\widehat{i}j}: 
\begin{cases}
        \sij \rightarrow \lambda \sij\\
        \sik \rightarrow \frac{1}{\omzj}(\sik+\sjk)\\
        \sjk \rightarrow -\frac{\zj}{\omzj}(\sik+\sjk)\\
        \sijk \rightarrow (\sik+\sjk)
\end{cases}
\end{equation}
where we only keep terms of order $\lambda^{-1}$. 
The corresponding up-projector is given by
\begin{equation}
\label{eq:CijupIF}
\PCupIF_{\widehat{i}j}: 
    \begin{cases}
        \frac{\zj}{\omzj}\rightarrow -\frac{\sjk}{\sijk}\\
        \frac{1}{\omzj}\rightarrow \frac{\sik}{\sijk}\\
        \frac{1}{\zj}\rightarrow -\frac{\sik}{\sjk}\\
        \zj\rightarrow -\frac{\sjk}{\sik+\sij}\\
        (\sik+\sjk)\rightarrow \sijk,
    \end{cases}
\end{equation}
where we have regulated any potential singularity when $\sik \to 0$ with $\sij$.  This occurs whenever the limit involves the $P_{q\bar{q}\leftarrow G}$ or $P_{gg\leftarrow G}$ splitting functions. In practice, this means that contributions like $1/s_{ij}s_{ik}$ are split across two separate antenna by partial fractioning, 
\begin{equation}
    \frac{1}{s_{ij}s_{ik}} \to 
    \frac{1}{s_{ij}(s_{ik}+s_{ij})}
    +
    \frac{1}{s_{ik}(s_{ik}+s_{ij})}.
\end{equation}

When the two final-state particles $j$ and $k$ are collinear, the corresponding projectors are 
\begin{align}
\label{CkjdownIF}
\PCdownIF_{kj} &\equiv  \PCdownFF_{kj},\\
\PCupIF_{kj} &\equiv  \PCupFF_{kj}.
\end{align}

\subsubsection*{(II) Spectator in initial state.}

When the spectator particle is in the initial state, the down-projector is the same,
\begin{equation}
\label{eq:CijdownII}
\PCdownII_{\widehat{i}j} = \PCdownIF_{\widehat{i}j}
\end{equation}
but because $\sik$ can never be small, we do not have to be concerned by the appearance of $1/\sik$ in the up-projector, and we define
\begin{equation}
\label{eq:CijupII}
\PCupII_{\widehat{i}j}: 
    \begin{cases}
        \omzj\rightarrow \frac{\sijk}{\sik}\\
        \zj\rightarrow -\frac{\sjk}{\sik}\\
        (\sik+\sjk)\rightarrow \sijk.
    \end{cases}
\end{equation}

The projectors when $j$ becomes collinear to the initial-state hard radiator particle $k$ are obtained by exchanging the roles of $i$ and $k$,
\begin{align}
 \PCdownII_{\widehat{k}j} &=  \PCdownII_{\widehat{i}j} \qquad i \leftrightarrow k,\\
 \PCupII_{\widehat{k}j} &=  \PCupII_{\widehat{i}j} \qquad i \leftrightarrow k.
\end{align}

\subsection{Algorithm for Initial-Final Antennae}
\label{IFalg}
At NLO, we want to construct the initial-final three-particle antenna.  There are two types - identity-preserving and identity-changing.

The identity-preserving antenna functions are denoted by $\X(\widehat{i}_{a}^h,j_{b},k_{c}^{h})$, where the particle types are denoted by $a$, $b$, and $c$, which carry four-momenta $\widehat{i}$, $j$, and $k$ respectively. Particles $a$ and $c$ should be hard, and the antenna functions must have the correct limits when particle $b$ is unresolved. 

The identity-preserving antenna functions are characterised by three limits: the $b$ soft limit, the initial-final collinear limit between particles $\hat{a}$ and $b$, and the final-final collinear limit between particles $c^{h}$ and $b$. 
We systematically start from the most singular limit, and build the list of target functions from single-soft and simple-collinear limits,
\begin{align}
\label{eq:limitsIF}
L_1(\widehat{i},j,k^h) &= S_b(\widehat{i},j,k^h),\\
L_2(\widehat{i},j,k^h) &= P_{ab \leftarrow a} ( \widehat{i},j), \\
L_3(\widehat{i},j,k^h) &= P_{cb} ( k^h,j).
\end{align}

From these limits, we can then construct the antenna by applying the algorithm in three steps
\begin{equation}
\label{eq:algorithmIF}
\begin{split}
&X_{3;1}^0 (\widehat{i},j,k^h) = 
\PSup_{j}L_{1}(\widehat{i},j,k^h)\\
&X_{3;2}^0  (\widehat{i},j,k^h) = \PCupIF_{\widehat{i}j}(L_2(\widehat{i},j,k^h)-\PCdownIF_{\widehat{i}j} X_{3;1}^0 (\widehat{i},j,k^h))
+X_{3;1}^0 (\widehat{i},j,k^h)\\
&X_{3;3}^0  (\widehat{i},j,k^h) = \PCupIF_{kj}(L_3(\widehat{i},j,k^h)-\PCdownIF_{kj}X_{3;2}^0 (\widehat{i},j,k^h))
+X_{3;2}^0 (\widehat{i},j,k^h)
\end{split}
\end{equation}
and then take
\begin{equation}
\X (\widehat{i},j,k^h) \equiv X_{3;3}^0 (\widehat{i},j,k^h).
\end{equation}
In particular, this guarantees that
\begin{equation}
\begin{split}
&\PSdown_{j}\X (\widehat{i},j,k^h) = L_1(\widehat{i},j,k^h),\\
&\PCdownIF_{\widehat{i}j}\X (\widehat{i},j,k^h) = L_{2}(\widehat{i},j,k^h),\\
&\PCdownIF_{kj}\X (\widehat{i},j,k^h) = L_{3}(\widehat{i},j,k^h).
\end{split}
\end{equation}
Similar equations apply to the identity-preserving $\X(k^h,j,\widehat{i})$ antennae.

The identity-changing antenna functions are denoted by 
$\X(j_{a},\widehat{i}_{b},k_{c}^{h})$, where the particle types $a$, $b$, and $c$ now carry four-momenta $j$, $\widehat{i}$, and $k$ respectively. $\X (j_{a},\widehat{i}_{b},k_{c}^{h})$ is characterised by only one non-zero limit: the initial-final collinear limit between particles $a$ and $\widehat{b}$. Therefore, 
\begin{align}
\label{eq:limitsIF-IC}
L_1(j,\widehat{i},k^h) &= P_{ba \leftarrow A} (\widehat{i},j),
\end{align}
and
\begin{equation}
\begin{split}
&\X (j,\widehat{i},k^{h}) \equiv 
\PCupIF_{\widehat{i}j}L_{1}(j,\widehat{i},k^h).
\end{split}
\end{equation}
Similar equations apply to the identity-changing $\X(k^h,\widehat{i},j)$ antennae.

We use the antenna mapping given in Ref.~\cite{Daleo:2006xa} to absorb the unresolved momentum $j$ into the residual on-shell hard radiators $\widehat{I}$ and $K$,
$$
\{\widehat{i}, j, k^h\} \to  \{\widehat{I}, K\}.
$$
The invariant mass of the antenna is $\sijk = s_{\widehat{I}K} \equiv -Q^2$. 
The identity-preserving and identity-changing antennae integrated over the Initial-Final antenna phase space~\cite{Daleo:2006xa,ALTARELLI1979461} are given respectively by,
\begin{align}
\label{eq:XIFint}
\calX(\scaleIF,\xi) &= 
\QQIF
\frac{e^{\e\gamma}}{2\Gamma(1-\e)}
\left(\frac{1-\xi}{\xi}\right)^{-\e}\int_{0}^{1} dy (1-y)^{-\e}y^{-\e}
 Q^{2} \X,
\end{align}
where
\begin{equation}
\xi=\frac{\sijk}{\sik+\sij},
\qquad 
y = \frac{s_{ik}}{s_{ik}+s_{ij}}
\end{equation}
and 
\begin{equation}
\sjk=\frac{Q^{2}(1-\xi)}{\xi},\qquad
\sij=-\frac{Q^{2}(1-y)}{\xi},\qquad
\sik=-\frac{Q^{2}y}{\xi}.
\end{equation}

\subsection{Algorithm for Initial-Initial Antennae}
\label{IIalg}

Initial-initial antennae follow a similar pattern with both identity-preserving and identity-changing antennae.

The identity-preserving antenna functions are denoted by $\X(\widehat{i}_{a},j_{b},\widehat{k}_{c})$, where the particle types are denoted by $a$, $b$, and $c$, which carry four-momenta $\widehat{i}$, $j$, and $\widehat{k}$ respectively. They are characterised by three limits: the $b$ soft limit, the initial-final collinear limit between particles $\widehat{a}$ and $b$, and the initial-final collinear limit between particles $\widehat{c}$ and $b$. 

We systematically start from the most singular limit, and build the list of target functions from single-soft and identity-preserving simple-collinear limits,
\begin{align}
\label{eq:limitsII}
L_1(\widehat{i},j,\widehat{k}) &= S_b(\widehat{i},j,\widehat{k}),\\
L_2(\widehat{i},j,\widehat{k}) &= P_{ab \leftarrow a} ( \widehat{i},j), \\
L_3(\widehat{i},j,\widehat{k}) &= P_{cb \leftarrow c} ( \widehat{k},j).
\end{align}

From these limits, we can then construct the antenna by applying the algorithm
\begin{equation}
\label{eq:algorithmII}
\begin{split}
&X_{3;1}^0 (\widehat{i},j,\widehat{k}) = \PSup_{j}L_{1}(\widehat{i},j,\widehat{k})\\
&X_{3;2}^0 (\widehat{i},j,\widehat{k}) = \PCupII_{\widehat{i}j}(L_2(\widehat{i},j,\widehat{k})-\PCdownII_{\widehat{i}j}X_{3;1}^0 (\widehat{i},j,\widehat{k}))+X_{3;1}^0 (\widehat{i},j,\widehat{k})\\
&X_{3;3}^0 (\widehat{i},j,\widehat{k}) = \PCupII_{\widehat{k}j}(L_3(\widehat{i},j,\widehat{k})-\PCdownII_{\widehat{k}j}X_{3;2}^0 (\widehat{i},j,\widehat{k}))+X_{3;2}^0 (\widehat{i},j,\widehat{k})
\end{split}
\end{equation}
and then
\begin{equation}
\X (\widehat{i},j,\widehat{k}) = X_{3;3}^0 (\widehat{i},j,\widehat{k}).
\end{equation}
In particular, this guarantees that
\begin{equation}
\begin{split}
&\PSdown_{j}\X (\widehat{i},j,\widehat{k}) = L_{1}(\widehat{i},j,\widehat{k})\\
&\PCdownII_{\widehat{i}j}\X (\widehat{i},j,\widehat{k}) = L_{2}(\widehat{i},j,\widehat{k})\\
&\PCdownII_{\widehat{k}j}\X (\widehat{i},j,\widehat{k}) = L_{3}(\widehat{i},j,\widehat{k}).
\end{split}
\end{equation}
Similar equations apply to the identity-preserving $\X(\widehat{k},j,\widehat{i})$ antennae.

The identity-changing initial-initial antenna functions are denoted by, 
$\X(\widehat{i}_{a},\widehat{k}_{b},j_{c})$, where the particle types $a$, $b$, and $c$ now carry four-momenta $\widehat{i}$, $\widehat{k}$, and $j$ respectively. $\X (\widehat{i}_{a},\widehat{k}_{b},j_{c})$ is characterised by only one non-zero limit: the initial-final collinear limit between particles $c$ and $\widehat{b}$. Therefore, 
\begin{align}
\label{eq:limitsII-IC}
L_1(\widehat{i},\widehat{k},j) &= P_{bc \leftarrow C} (\widehat{k},j),
\end{align}
and
\begin{equation}
\begin{split}
&\X (\widehat{i},\widehat{k},j) \equiv 
\PCupII_{\widehat{k}j}L_{1}(\widehat{i},\widehat{k},j).
\end{split}
\end{equation}
Similar equations apply to the identity-changing $\X(j,\widehat{k},\widehat{i})$ antennae.

We use the antenna mapping given in Ref.~\cite{Daleo:2006xa} to absorb the unresolved momentum $j$ into the residual on-shell hard radiators $\widehat{I}$ and $\widehat{K}$,
$$
\{\widehat{i}, j, \widehat{k}\} \to  \{\widehat{I}, \widehat{K}\}.
$$
The invariant mass of the antenna is $\sijk = s_{\widehat{I}\widehat{K}} \equiv Q^2$. 
The identity-preserving and identity-changing antennae integrated over the Initial-Initial antenna phase space~\cite{Daleo:2006xa} are given by
\begin{align}
\label{eq:XIIint}
\calX(\scaleII,\xi,\xk)&=
\QQII
\frac{e^{\e\gamma}}{\Gamma(1-\e)}\mathcal{J}(\xi,\xk)\, Q^{2}\, \X,
\end{align}
where
\begin{align}
\mathcal{J}(\xi,\xk)&=\frac{\xi\xk(1+\xi\xk)}{(\xi+\xk)^{2}}(1-\xi)^{-\e}(1-\xk)^{-\e}
\left(\frac{(1+\xi)(1+\xk)}{(\xi+\xk)^{2}}\right)^{-\e}
\end{align}
with
\NG{
\begin{align}
    \xi = \sqrt{\frac{Q^2 (\sik+\sjk) }{\sik (\sij+\sik)}}, 
\qquad     
    \xk = \sqrt{\frac{Q^2 (\sij+\sik) }{\sik (\sik+\sjk)}}, 
\end{align}
}
and 
\begin{equation}
\sijk = Q^{2},\qquad \sik =\frac{Q^2}{\xi\xk}, \qquad
\sij = -Q^2\frac{(1-\xk^{2})}{\xk(\xi+\xk)}, \qquad
\sjk = -Q^2\frac{(1-\xi^{2})}{\xi(\xi+\xk)}.
\end{equation}

\section{Initial-Final Antennae}
\label{sec:IF}

In this section, we apply the algorithm outlined in Section~\ref{IFalg} to construct $\X$ antennae with initial-final kinematics from the relevant limits and compare them with the antennae derived from matrix elements given in Ref.~\cite{Daleo:2006xa}, denoted by $\Xold$. We expect that the new $\X$ antennae will only differ from the $\Xold$ antenna by terms that are not singular at any point in phase space, or by terms that vanish as $\e \to 0$. In this case, we expect that the integrated antennae $\calX$ differ from the corresponding integrated antenna $\calXold$ by terms of $\order{\e^0}$ and/or by terms that are regular as $\xi \to 1$ (i.e. not distributions).

As indicated in Table~\ref{tab:X30IF}, there are six distinct IP initial-final antennae and four distinct IC antennae, each accounting for two particle assignments. We note that each Final-Final antenna configuration gives rise to four Initial-Final antennae - two of Type 1 and two of Type 2.  Therefore, the eight Final-Final configurations listed in Table~\ref{tab:X30FF} give rise to thirty-two Initial-Final configurations.  Twenty IF configurations are listed in Table~\ref{tab:X30IF}.  The remaining twelve configurations are not needed.   They fall into three classes: 
\begin{enumerate}

\item 
$\X(g^h,\widehat{g},g),\qquad
\X(g,\widehat{g},g^h),\qquad
\X(q^h,\widehat{g},g),\qquad
\X(g,\widehat{g},\bar{q}^h)$

The $\widehat{g}||g$ collinear limits are fully contained in the $\gF$ and $\gD$ antennae.

\item
$\X(q^h,\qpbar,\widehat{\qp}),\qquad
\X(\widehat{\qpbar},\qp,\bar{q}^h),\qquad
\X(g^h,\qpbar,\widehat{\qp}),\qquad
\X(\widehat{\qpbar},\qp,g^h)$

These are IC configurations describing the $\qp\qpbar$ collinear limit.  This limit is entirely described by the $\QtogE$ and $\QtogG$ antennae.

\item 
$\X(q,\widehat{\qpbar},\qp^h),\qquad
\X(\qpbar^h,\widehat{\qp},\bar{q}),\qquad
\X(g,\widehat{\qpbar},\qp^h),\qquad
\X(\qpbar^h,\widehat{\qp},g)$

The two hard radiators can never be collinear, so these antennae vanish.   
\end{enumerate}

When the antenna is built from limits that are simply obtained by crossing, then we expect that the antenna is also obtained by crossing.  This is not always the case for IC antenna (which only contain one limit) or for IP antenna which contain two collinear gluons.  As discussed earlier, the full final-final $g||g$ collinear limit is obtained by combining two antenna, one with the $g^h||g$ limit and one with the $g||g^h$ limit.  However, the $\widehat{g}||g$ collinear limit is contained in a single antenna.  Antennae where one of the two colour-connected gluons is crossed to the initial state are therefore not obtained by crossing.


\subsection{Identity-preserving initial-final antennae}

As indicated in Table~\ref{tab:X30IF}, there are six identity-preserving antenna: three quark-initiated and three gluon-initiated.

\subsubsection{$\qA(\widehat{i}_{q},j_{g},k_{\bar{q}}^{h})$}
Bulding the antenna iteratively according to Eq.~\eqref{eq:algorithmIF} using the list of limits in Eq.~\eqref{eq:limitsIF}, we find that the three-parton tree-level antenna function with quark-antiquark parents is given (to all orders in $\e$) by
\begin{equation}
\label{eq:qA30IF}
\qA(\widehat{i}_{q},j_{g},k_{\bar{q}}^{h}) = \frac{2\sik}{\sij\sjk}+\frac{\sij\ome}{\sjk\sijk}+\frac{\sjk\ome}{\sij\sijk}.
\end{equation}
Comparing our result to the corresponding antenna function $\qAold$ derived from matrix elements given in Ref.~\cite{Daleo:2006xa}, we find agreement to $\order{\e}$,
\begin{equation}
\qAold(\widehat{i}_{q},j_{g},k_{q}) = \qA(\widehat{i}_{q},j_{g},k_{\bar{q}}^{h})+\order{\e}.
\end{equation}
We also observe that Eq.~\eqref{eq:qA30IF} can also be obtained by crossing the $\A$ antenna for final-final kinematics given in Ref.~\cite{paper2}.

Integrating Eq.~\eqref{eq:qA30IF} over the initial-final antenna phase space yields,
\begin{equation}
\label{eq:calqA30IF}
\begin{split}
\calqA(\scaleIF,\xi)&=
-2\bold{I}_{q\bar{q}}^{(1)}(\e,\scaleIF)\delta(1-\xi)
+\QQIF\Big(-\frac{1}{2\e}p_{qq}^{(0)}(\xi)
\\
&+\left(\frac{7}{4}-\frac{\pi^{2}}{6}\right)\delta(1-\xi)
-\frac{3}{4}\Dox+\Dlx+\frac{3-\xi}{2}\\
&-\frac{1+\xi}{2}\log(1-\xi)
-\frac{1+\xi^{2}}{2(1-\xi)}\log(\xi)\Big)+\order{\e}.
\end{split}
\end{equation}
where $\bold{I}_{q\bar{q}}^{(1)}$ is the Catani infrared singularity operator listed in Appendix~\ref{sec:catani}, $p_{qq}^{(0)}(x)$ is the colour ordered splitting kernel given in Appendix~\ref{sec:kernels}, and we have introduced the distributions
\begin{equation}
\mathcal{D}_{n}(x)=\left(\frac{\log^{n}(1-x)}{1-x}\right)_{+}.
\end{equation}
Once we correct for the typo (flipped sign on the non-singular part) in Ref.~\cite{Daleo:2006xa}, this agrees with the expression for $\calqAold$ up to $\order{\e}$. 

\subsubsection{$\qD(\widehat{i}_{q},j_{g},k_{g}^{h})$}
The $\qD(\widehat{i}_{q},j_{g},k_{g}^{h})$ antenna has singular limits when the unresolved gluon becomes soft, or becomes collinear to either the initial-state hard radiator quark or final-state hard radiator gluon.
Using the algorithm, we obtain the antenna
\begin{equation}
\label{eq:qD30IF}
\qD(\widehat{i}_{q},j_{g},k_{g}^{h})= \frac{2\sik}{\sij\sjk}
+\frac{\ome\sjk}{\sij\sijk}+\frac{\sij\sik}{\sjk\sijk^{2}}.
\end{equation}
We observe that Eq.~\eqref{eq:qD30IF} can also be obtained by crossing the $\D$ antenna for final-final kinematics given in Ref.~\cite{paper2}.
Compared to the antenna derived from the matrix elements for neutralino decay in Ref.~\cite{Daleo:2006xa} where either gluon could be soft, we find that  
\begin{equation}
\begin{split}
\qDold(\widehat{i}_{q},j_{g},k_{g})&=\qD(\widehat{i}_{q},j_{g},k_{g}^{h})
+\qD(\widehat{i}_{q},k_{g},j_{g}^{h})\\
&+\frac{4\sjk}{\sijk^{2}}+\frac{5\sij}{\sijk^{2}}+\frac{5\sik}{\sijk^{2}} +\order{\e}.
\end{split}
\end{equation}
Apart from the doubling up of antenna due to the requirement that one of the final-state gluons is hard, the remaining terms like $\sij/\sijk$ are not singular at any point in phase space. 

Integrating Eq.~\eqref{eq:qD30IF} over the initial-final antenna phase space yields
\begin{equation}
\label{eq:calqD30IF}
\begin{split}
\calqD(\scaleIF,\xi)
&=-2\bold{I}_{qg}^{(1)}(\e,\scaleIF)\delta(1-\xi)+\QQIF\Big(-\frac{1}{2\e}p_{qq}^{(0)}(\xi)\\
&+\left(\frac{67}{36}-\frac{\pi^{2}}{6}\right)\delta(1-\xi)-\frac{11}{12}\Dox+\Dlx+\frac{1}{12\xi}+\frac{3-\xi}{2}\\
&-\frac{1+\xi}{2}\log(1-\xi)
-\frac{1+\xi^{2}}{2(1-\xi)}\log(\xi)\Big)+\order{\e}.
\end{split}
\end{equation}
As expected, this agrees with half of the expression in Ref.~\cite{Daleo:2006xa} for $\calqDold$ up to non-singular terms (i.e. terms that are regular as $\xi \to 1$) at $\order{\e^0}$.

\subsubsection{$\qE(\widehat{i}_{q},j_{\qpbar},k_{\qp}^{h})$}
The quark-initiated three quark antenna $\qE(\widehat{i}_{q},j_{\qpbar},k_{\qp}^{h})$ has no soft limit as the unresolved particle is a quark, but does have a limit when that quark becomes collinear to the final-state hard anti-quark of the same type. There is no corresponding collinear singularity with the initial-state quark. 
Using the algorithm, we obtain the antenna
\begin{equation}
\label{eq:qE30IF}
    \qE(\widehat{i}_{q},j_{\qpbar},k_{\qp}^{h}) = \frac{1}{\sjk}-\frac{2\sij\sik}{\sjk\sijk^{2}\ome}.
\end{equation}
We observe that Eq.~\eqref{eq:qE30IF} can also be obtained by crossing the $\E$ antenna for final-final kinematics given in Ref.~\cite{paper2}.
Compared to the antenna derived from matrix elements given in Ref.~\cite{Daleo:2006xa}, we find that  
\begin{equation}
    \qEold(\widehat{i}_{q},j_{\qpbar},k_{\qp}) = \qE(\widehat{i}_{q},j_{\qpbar},k_{\qp}^{h})
    -\frac{\sjk}{\sijk^{2}}-\frac{\sik}{\sijk^{2}}-\frac{\sij}{\sijk^{2}}+\order{\e}.
\end{equation}
We find agreement up to $\order{\e}$ plus terms that are 
not singular at any point in phase space such as $\sij/\sijk^{2}$.

Integrating Eq.~\eqref{eq:qE30IF} over the initial-final antenna phase space yields
\begin{equation}
\label{eq:calqE30IF}
\begin{split}
\calqE(\scaleIF,\xi)
&=-4\bold{I}_{qg,F}^{(1)}(\e,\scaleIF)\delta(1-\xi)\\
&+\QQIF\Big(-\frac{5}{9}\delta(1-\xi)+\frac{1}{3}\Dox-\frac{1}{6\xi}-\frac{1}{2}\Big)+\order{\e},
\end{split}
\end{equation}
which agrees with the expression in Ref.~\cite{Daleo:2006xa} for $\calqEold$ (after correcting a typo ${\mathcal D}_1 \to {\mathcal D}_0$) up to non-singular terms at $\order{\e^0}$.

\subsubsection{$\gD(k_{q}^{h},j_{g},\widehat{i}_{g})$}
For the gluon-initiated $qgg$ antenna, $\gD(k_{q}^{h},j_{g},\widehat{i}_{g})$, there are limits when gluon $j$ goes collinear to either the initial-state hard gluon or the final-state hard quark.
We find
\begin{equation}
\label{eq:gD30IF}
\gD(k_{q}^{h},j_{g},\widehat{i}_{g}) = \frac{2\sik}{\sij\sjk}+\frac{2\sik\sjk}{\sijk^{2}\sij}+\frac{\sij\ome}{\sjk\sijk}+\frac{2\sjk}{\sij(\sik+\sij)}.
\end{equation}
This is the first antenna involving the $P_{gg\leftarrow G}$ splitting function, and the term proportional to the momentum fraction $\zj$ produces the term with $(\sik+\sij)$ in the denominator.
Comparing with the antenna derived from matrix elements given in Ref.~\cite{Daleo:2006xa}, we find that 
\begin{equation}
\begin{split}
\gDold(k_{q},j_{g},\widehat{i}_{g})&=
\gD(k_{q}^{h},j_{g},\widehat{i}_{g})
+\frac{7\sjk}{\sijk^2}+\frac{5\sij}{\sijk^2}
+\frac{5\sik}{\sijk^2} +\order{\e}
\end{split}
\end{equation}
and observe that agree up to terms of $\order{\e}$ and terms that are not singular over the whole of phase space.
Note that this antenna cannot be obtained by crossing the $\D$ antenna for final-final kinematics given in Ref.~\cite{paper2} because this antenna contains the full $\widehat{i}_{g},j_{g}$ collinear limit. Note also the presence of the $\sik+\sij$ denominator.  This is to avoid potential singularities as $\sik \to 0$ and is generated by partial fractioning.  The partner term appears in the flavour-changing $\gtoqD(j_{q},\widehat{i}_{g},k_{g}^h)$ antenna discussed in section~\ref{subsec:gtoqD}. A similar split was also performed in Ref.~\cite{Daleo:2006xa}.

Integrating Eq.~\eqref{eq:gD30IF} over the initial-final antenna phase space yields
\begin{equation}
\label{eq:calgD30IF}
\begin{split}
\calgD(\scaleIF,\xi)&=-2\bold{I}^{(1)}_{qg}(\e,\scaleIF)\delta(1-\xi)+\QQIF\Big(-\frac{1}{2\e}p_{gg}^{(0)}(\xi)\\
&+2-\frac{1}{\xi}+\Dlx-\frac{3}{4}\Dox+\left(\frac{7}{4}-\frac{\pi^{2}}{6}\right)\delta(1-\xi)\\
&-\frac{(1-\xi+\xi^2)^{2}}{\xi(1-\xi)}\log(\xi)+\frac{1-2\xi+\xi^{2}-\xi^{3}}{\xi}\log(1-\xi)\Big)+\order{\e},
\end{split}
\end{equation}
which agrees with the expression in Ref.~\cite{Daleo:2006xa} up to non-singular terms at $\order{\e^0}$.

\subsubsection{$\gF(\widehat{i}_{g},j_{g},k_{g}^{h})$}
For the three gluon antenna $\gF(\widehat{i}_{g}^{h},j_{g},k_{g}^{h})$, we have limits when the unresolved gluon goes soft, and when it goes collinear with either the initial-state gluon or final-state hard gluon.
We find
\begin{equation}
\label{eq:gF30IF}
\gF(\widehat{i}_{g},j_{g},k_{g}^{h})= \frac{2\sik}{\sij\sjk}+\frac{2\sik\sjk}{\sijk^{2}\sij}+\frac{\sij\sik}{\sijk^{2}\sjk}+\frac{2\sjk}{\sij(\sik+\sij)}
\end{equation}
once again reflecting the presence of the $P_{gg\leftarrow G}$ splitting function. Note that this antenna cannot be obtained by crossing the $\F$ antenna for final-final kinematics given in Ref.~\cite{paper2} because this antenna contains the full $\widehat{i}_{g},j_{g}$ collinear limit.
Compared to the matrix-element derived antenna given in  Ref.~\cite{Daleo:2006xa}, where either of the final-state gluons could be soft, we find that 
\begin{equation}
\begin{split}
\gFold(\widehat{i}_{g},j_{g},k_{g}) &= \gF(\widehat{i}_{g},j_{g},k_{g}^{h}) + \gF(\widehat{i}_{g},k_{g},j_{g}^{h})\\
&+ \frac{8\sjk}{\sijk^{2}} + \frac{8\sij}{\sijk^{2}} + \frac{8\sik}{\sijk^{2}} + \order{\e}.
\end{split}
\end{equation}
The differences are finite when $\e \to 0$ and are not singular anywhere in phase space.

Integrating Eq.~\eqref{eq:gF30IF} over the initial-final antenna phase space yields,
\begin{equation}
\begin{split}
\label{eq:calgF30IF}
\calgF(\scaleIF,\xi)&=-2\bold{I}^{(1)}_{gg}(\e,\scaleIF)\delta(1-\xi)+\QQIF\Big(-\frac{1}{2\e}p_{gg}^{(0)}(\xi)\\
&+\left(\frac{67}{36}-\frac{\pi^{2}}{6}\right)\delta(1-\xi)+\Dlx-\frac{11}{12}\Dox+2-\frac{11}{12\xi}\\
&+\frac{(1-2\xi+\xi^2-\xi^3)}{\xi}\log(1-\xi)-\frac{(1-\xi+\xi^{2})^{2}}{\xi(1-\xi)}\log(\xi)\Big)+\order{\e}.
\end{split}
\end{equation}
As expected, this agrees with half of the expression in Ref.~\cite{Daleo:2006xa} for $\calgFold$ up to non-singular terms at $\order{\e^0}$.

\subsubsection{$\gG(\widehat{i}_{g},j_{\qpbar},k_{\qp}^{h})$}
The gluon-initiated antenna function $\gG(\widehat{i}_{g},j_{\qpbar},k_{\qp}^{h})$ has the same limits as the quark-initiated antenna $\qE(\widehat{i}_{q}^{h},j_{\qpbar},k_{\qp}^{h})$, which means using the algorithm we obtain the same result for both. 
We find
\begin{equation}
\label{eq:gG30IF}
    \gG(\widehat{i}_{g},j_{\qpbar},k_{\qp}^{h}) = \frac{1}{\sjk}-\frac{2\sij\sik}{\sjk\sijk^{2}\ome}.
\end{equation}
We observe that Eq.~\eqref{eq:gG30IF} can also be obtained by crossing the $\G$ antenna for final-final kinematics given in Ref.~\cite{paper2}.
The antenna derived from matrix elements~\cite{Daleo:2006xa} is related to $\gG$ by
\begin{equation}
\gGold(\widehat{i}_{g},j_{\qpbar},k_{\qp})= \gG(\widehat{i}_{g},j_{\qpbar},k_{\qp}^{h}) -\frac{\sjk}{\sijk^{2}}-\frac{2\sik}{\sijk^{2}}-\frac{2\sij}{\sijk^{2}}+\order{\e}
\end{equation}
and we see they agree up to non-singular terms at $\order{\e^0}$ as we would expect.

Integrating Eq.~\eqref{eq:gF30IF} over the initial-final antenna phase space yields,
\begin{equation}
\begin{split}
\label{eq:calgG30IF}
\calgG(\scaleIF,\xi)
&=-4\bold{I}_{qg,F}^{(1)}(\e,\scaleIF)\delta(1-\xi)
\\
&+\QQIF\Big(-\frac{5}{9}\delta(1-\xi)+\frac{1}{3}\Dox
-\frac{1}{6\xi}-\frac{1}{2}\Big)+\order{\e}
\end{split}
\end{equation}
which agrees with Eq.~\eqref{eq:calqE30IF}.
After correcting for typos, this also agrees with
the expression in Ref.~\cite{Daleo:2006xa} for $\calgGold$ up to non-singular terms at $\order{\e^0}$.

\subsection{Identity-changing initial-final antennae}

As indicated in Table~\ref{tab:X30IF}, there are four identity-changing antenna: two describing gluon to quark transitions and two describing quark to gluon transitions.  

\subsubsection{$\gtoqA(j_{q},\widehat{i}_{g},k_{\bar{q}}^h)$}
For the identity-changing $\gtoqA(j_{q},\widehat{i}_{g},k_{\bar{q}}^h)$ antenna, the gluon is in the initial state and when it becomes collinear with the final-state quark, the identity of the initial state changes. This is the only limit in this antenna, and is controlled by the $P_{q\bar{q}\leftarrow G}$ splitting function where the term proportional to the momentum fraction $\zj$ produces a term in the antenna with $(\sik+\sij)$ in the denominator. 
We find
\begin{equation}
\label{eq:gtoqA30IF}
\gtoqA(j_{q},\widehat{i}_{g},k_{\bar{q}}^h)= -\frac{\ome\sik}{\sij\sijk}-\frac{2\sjk}{\sij(\sik+\sij)}.
\end{equation}
The antenna derived from matrix elements in Ref.~\cite{Daleo:2006xa} contains limits when the initial-state gluon is collinear with the final-state quark and the final-state anti-quark.  Therefore, we find that 
\begin{equation}
\gtoqAold(j_{q},\widehat{i}_{g},k_{\bar{q}})= 
\gtoqA(j_{q},\widehat{i}_{g},k_{\bar{q}}^h)+
\gtoqA(k_{\bar{q}},\widehat{i}_{g},j_{q}^h) +\order{\e}.
\end{equation}
Integrating Eq.~\eqref{eq:gtoqA30IF} over the initial-final antenna phase space yields
\begin{equation}
\label{eq:calgtoqA30IF}
\calgtoqA(\scaleIF,\xi)=\QQIF\Big(-\frac{1}{2\e}p_{qg}^{(0)}(\xi)-\left(\frac{1}{2}-\xi+\xi^{2}\right)\log\left(\frac{\xi}{1-\xi}\right)\Big)+\order{\e}.
\end{equation}
After correcting for typos, this agrees with half of the expression in Ref.~\cite{Daleo:2006xa} for $\calgtoqAold$ up to non-singular terms  at $\order{\e^0}$.

\subsubsection{$\gtoqD(j_{q},\widehat{i}_{g},k_{g}^h)$}
\label{subsec:gtoqD}
The identity-changing antenna function $\gtoqD(j_{q},\widehat{i}_{g},k_{g}^h)$ only has one collinear limit, when the final-state quark is collinear with the initial-state gluon.  This is the same limit as the $\gtoqA(j_{q},\widehat{i}_{g},k_{\bar{q}}^h)$ antenna,
and therefore the algorithm produces the same result,
\begin{equation}
\label{eq:gtoqD30IF}
\gtoqD(j_{q},\widehat{i}_{g},k_{g}^h)= -\frac{\ome\sik}{\sij\sijk}-\frac{2\sjk}{\sij(\sik+\sij)}.
\end{equation}
This agrees with the antenna given in Ref.~\cite{Daleo:2006xa} up to non-singular terms,
\begin{equation}
\gtoqDold(j_{q},\widehat{i}_{g},k_{g}) = \gtoqD(j_{q},\widehat{i}_{g},k_{g}^h)+\frac{\sik}{\sijk^{2}}+\frac{2\sjk}{\sijk^{2}}+\order{\e}.
\end{equation}

Integrating Eq.~\eqref{eq:gtoqD30IF} over the initial-final antenna phase space yields
\begin{equation}
\label{eq:calgtoqD30IF}
\calgtoqD(\scaleIF,\xi)=
\QQIF\Big(-\frac{1}{2\e}p_{qg}^{(0)}(\xi)-\left(\frac{1}{2}-\xi+\xi^{2}\right)\log\left(\frac{\xi}{1-\xi}\right)\Big)+\order{\e},
\end{equation}
which agrees with the expression in Ref.~\cite{Daleo:2006xa} for $\calgtoqDold$ up to non-singular terms  at $\order{\e^0}$.

\subsubsection{$\QtogE(k_{q}^h,\widehat{i}_{\qpbar},j_{\qp})$}
For the $\QtogE(k_{q}^h,\widehat{i}_{\qpbar},j_{\qp})$ identity-changing antenna, there is only one limit when the quarks of the same identity become collinear and form a gluon.
Using the algorithm, we obtain
\begin{equation}
\label{eq:QtogE30IF}
\QtogE(k_{q}^h,\widehat{i}_{\qpbar},j_{\qp}) = -\frac{1}{\sij}+\frac{2\sik\sjk}{\sij\sijk^{2}\ome}
\end{equation}
which agrees up to $\order{\e}$ and non singular terms with the antenna given in Ref.~\cite{Daleo:2006xa}.
The same result could be obtained by crossing the $\E(k,i,j)$ antenna for final-final kinematics given in Ref.~\cite{paper2}.
\begin{equation}
\QtogEold(k_{q},\widehat{i}_{\qpbar},j_{\qp}) = \QtogE(k_{q}^h,\widehat{i}_{\qpbar},j_{\qp})+\frac{\sjk}{\sijk^{2}}+\frac{\sik}{\sijk^{2}}+\frac{\sij}{\sijk^{2}}+\order{\e}.
\end{equation}

Integrating Eq.~\eqref{eq:QtogE30IF} over the initial-final antenna phase space yields
\begin{equation}
\begin{split}
\calQtogE(\scaleIF,\xi)
&=\QQIF\Big(-\frac{1}{2\e}p_{gq}^{(0)}(\xi)+2-\frac{2}{\xi}+\frac{(2-2\xi+\xi^{2})}{2\xi}\log\left(\frac{1-\xi}{\xi}\right)\Big)\\
&+\order{\e},
\end{split}
\end{equation}
which, after fixing typos, agrees with the result for $\calQtogEold$ given in Ref.~\cite{Daleo:2006xa} up to non-singular terms at $\order{\e^0}$.

\subsubsection{$\QtogG(k_{g}^h,\widehat{i}_{\qpbar},j_{\qp})$}
The $\QtogG(k_{g}^h,\widehat{i}_{\qpbar},j_{\qp})$ identity-changing antenna function has the same limit when the two same flavour quarks become collinear as the $\QtogE(k_{q}^h,\widehat{i}_{\qpbar},j_{\qp})$ antenna.  
Therefore, we find that the two antennae are equivalent,
\begin{equation}
\label{eq:QtogG30IF}
\QtogG(k_{g}^h,\widehat{i}_{\qpbar},j_{\qp})= -\frac{1}{\sij}+\frac{2\sik\sjk}{\sij\sijk^{2}\ome}.
\end{equation}
The same result could be obtained by crossing the $\G$ antenna for final-final kinematics given in Ref.~\cite{paper2}.

Comparing with the antenna given in Ref.~\cite{Daleo:2006xa},
\begin{equation}
\QtogGold(k_{g},\widehat{i}_{\qpbar},j_{\qp}) = \QtogG(k_{g}^h,\widehat{i}_{\qpbar},j_{\qp})+\frac{\sij}{\sijk^{2}}+\frac{2\sik}{\sijk^{2}}+\frac{2\sjk}{\sijk^{2}}+\order{\e}
\end{equation}
we see that they agree up to up to $\order{\e}$ and non singular terms.

Integrating Eq.~\eqref{eq:QtogG30IF} over the initial-final antenna phase space yields
\begin{equation}
\begin{split}
\calQtogG(\scaleIF,\xi)
&=\QQIF\Big(-\frac{1}{2\e}p_{gq}^{(0)}(\xi)+2-\frac{2}{\xi}+\frac{(2-2\xi+\xi^{2})}{2\xi}\log\left(\frac{1-\xi}{\xi}\right)\Big)\\
&+\order{\e},
\end{split}
\end{equation}
which agrees with the result for $\calQtogGold$ given in Ref.~\cite{Daleo:2006xa} up to non-singular terms  at $\order{\e^0}$.

\section{Initial-Initial Antennae}
\label{sec:II}
In this section, we apply the algorithm outlined in Section~\ref{IIalg} to construct $\X$ antennae with initial-initial kinematics from the relevant limits and compare them with the antennae derived from matrix elements given in Ref.~\cite{Daleo:2006xa}, denoted by $\Xold$. We expect that the new $\X$ antennae will only differ from the $\Xold$ antenna by terms that are not singular at any point in phase space, or by terms that vanish as $\e \to 0$. In this case, we expect that the integrated antennae $\calX$ differ from the corresponding integrated antenna $\calXold$ by terms of $\order{\e^0}$.

As indicated in Table~\ref{tab:X30II}, there are three distinct IP initial-final antennae and four distinct IC antennae. We note that each Final-Final antenna configuration gives rise to three Initial-Initial antennae - one of Type 1 and two of Type 2.  Therefore, the eight Final-Final configurations listed in Table~\ref{tab:X30FF} give rise to twenty-four Initial-Initial configurations.  Twelve configurations are listed in Table~\ref{tab:X30II}.  The remaining twelve configurations are not needed.   They fall into three classes: 
\begin{enumerate}

\item 
$\X(\widehat{g},\widehat{g},g),\qquad
\X(g,\widehat{g},\widehat{g}),\qquad
\X(\widehat{q},\widehat{g},g),\qquad
\X(g,\widehat{g},\widehat{\bar{q}})$

The $\widehat{g}||g$ collinear limits are fully contained in the $\ggF$ and $\qgD$ antennae.

\item
$\X(\widehat{q},\qpbar,\widehat{\qp}),\qquad
\X(\widehat{\qpbar},\qp,\widehat{\bar{q}}),\qquad
\X(\widehat{g},\qpbar,\widehat{\qp}),\qquad
\X(\widehat{\qpbar},\qp,\widehat{g})$

These are IC configurations describing the $\qp\qpbar$ collinear limit.  This limit is entirely described by the $\qQE$ and $\gQG$ antennae.

\item 
$\X(q,\widehat{\qpbar},\widehat{\qp}),\qquad
\X(\widehat{\qpbar},\widehat{\qp},\bar{q}),\qquad
\X(g,\widehat{\qpbar},\widehat{\qp}),\qquad
\X(\widehat{\qpbar},\widehat{\qp},g)$

The two hard radiators can never be collinear, so these antenna vanish.   
\end{enumerate}

As in the Initial-Final case, when the antenna is built from limits that are simply obtained by crossing, then we expect that the antenna is also obtained by crossing.  This is not always the case for IC antenna (which only have one limit) or for IP antennae which contain a collinear $g^h||g$ limit.  As discussed earlier, the full final-final $g||g$ collinear limit is obtained by combining two antenna, one with the $g^h||g$ limit and one with the $g||g^h$ limit.  However, the $\widehat{g}||g$ collinear limit is contained in a single antenna.  Antennae containing partonic configurations where one of the two colour connected gluons is crossed to the initial state are therefore not obtained by crossing. 

\subsection{Identity-preserving initial-initial antennae}

As indicated in Table~\ref{tab:X30II}, there are three identity-preserving antenna.

\subsubsection{$\qqA(\widehat{i}_{q},j_{g},\widehat{k}_{\bar{q}})$}

The $qg\bar{q}$ antenna with the quark and antiquark in the initial state has limits when the gluon is soft, or collinear with either of the initial-state particles. 
Using the algorithm, we obtain 
\begin{equation}
\label{eq:qqA30II}
\qqA(\widehat{i}_{q},j_{g},\widehat{k}_{\bar{q}})= 
\frac{2\sik}{\sij\sjk}
+\frac{\sjk\ome}{\sij\sijk}
+\frac{\sij\ome}{\sjk\sijk}
\end{equation}
which agrees with the antenna given in Ref.~\cite{Daleo:2006xa} up to $\order{\e}$,
\begin{equation}
\qqAold(\widehat{i}_{q},j_{g},\widehat{k}_{\bar{q}})= \qqA(\widehat{i}_{q},j_{g},\widehat{k}_{\bar{q}})+\order{\e}.
\end{equation}
We also observe that Eq.~\eqref{eq:qqA30II} can also be obtained by crossing the $\A$ antenna for final-final kinematics given in Ref.~\cite{paper2}.

Integrating Eq.~\eqref{eq:qqA30II} over the initial-initial antenna phase space yields
\begin{equation}
\label{eq:calqqA}
\begin{split}
&\calqqA(\scaleII,\xi,\xk) = -\bold{I}_{q\bar{q}}^{(1)}(\e,\scaleII)\delta(1-x_{i})\delta(1-x_{k})+\QQII\Big(-\frac{1}{2\e}p_{qq}^{(0)}(x_{i})\delta(1-x_{k})\\
&+\delta(1-x_{i})\Dlxii+\frac{\pi^{2}}{4}\delta(1-x_{i})\delta(1-x_{k})+\frac{1}{2}\Doxi\Doxii-\frac{1+x_{k}}{2}\Doxi\\
&+\left(\frac{(1-x_{i})^{2}+(1-x_{i})^{2}\log(\frac{2}{1-x_{i}^{2}})-2x_{i}^{2}\log(\frac{1+x_{i}}{2})}{2(1-x_{i})}\right)\delta(1-x_{k})\\
&+\frac{Y_1(x_{i},x_{k})-Y_1(x_{i},1)-Y_1(1,x_{k})+Y_1(1,1)}{2(1-x_{i})(1-x_{k})}+ (x_{i}\leftrightarrow x_{k})\Big)+\order{\e},
\end{split}
\end{equation}
where
\begin{equation}
Y_1(x_{i},x_{k})=\frac{(1+x_{i}x_{k})(2x_{i}x_{k}(x_{i}+x_{k})^{2}+x_{k}^{2}(1-x_{i}^{2})^{2}+x_{i}^{2}(1-x_{k}^{2})^{2})}{(x_{i}+x_{k})^{2}(1+x_{i})(1+x_{k})}.
\end{equation}
Note that the last term in Eq.~\ref{eq:calqqA} is clearly regular as $\xi \to 1$ or $\xk \to 1$.  We also note that Eq.~\ref{eq:calqqA} agrees with the result given in Ref.~\cite{Daleo:2006xa} to $\order{\e}$.

\subsubsection{$\qgD(\widehat{i}_{q},j_{g},\widehat{k}_{g})$}

The $qgg$ antenna with the quark and the colour-unconnected gluon ($k$) in the initial state has limits when the colour connected gluon ($j$) is soft, or collinear with either of the initial-state particles. 
Applying the algorithm yields
\begin{equation}
\label{eq:qgD30II}
\qgD(\widehat{i}_{q},j_{g},\widehat{k}_{g})= 
\frac{2\sik}{\sij\sjk}
+\frac{\sjk\ome}{\sij\sijk}
+\frac{2\sij}{\sik\sjk}
+\frac{2\sij\sik}{\sjk\sijk^{2}}.
\end{equation}
This antenna cannot be obtained by crossing the $\D$ antenna for final-final kinematics given in Ref.~\cite{paper2} because it includes the full $\widehat{g}g$ collinear limit, rather than the $g^hg$ collinear limit.

Comparing with the antenna given in Ref.~\cite{Daleo:2006xa}
\begin{equation}
\qgDold(\widehat{i}_{q},j_{g},\widehat{k}_{g}) =\qgD(\widehat{i}_{q},j_{g},\widehat{k}_{g})+\frac{\sjk}{\sik\sijk}+\frac{4\sjk}{\sijk^{2}}+\frac{5\sij}{\sijk^{2}}+\frac{5\sik}{\sijk^{2}}+\order{\e},
\end{equation}
we see that they agree up to $\order{\e}$ and non-singular terms. 

Integrating Eq.~\eqref{eq:qgD30II} over the initial-initial antenna phase space yields
\begin{equation}
\begin{split}
\calqgD(\scaleII,\xi,\xk)&=-2\bold{I}_{qg}^{(1)}(\e,\scaleII)\delta(1-x_{i})\delta(1-x_{k})+\QQII\Big(-\frac{1}{2\e}p_{qq}^{(0)}(x_{i})\delta(1-x_{k})\\
&-\frac{1}{2\e}p_{gg}^{(0)}(x_{k})\delta(1-x_{i})+\frac{\pi^{2}}{2}\delta(1-x_{i})\delta(1-x_{k})+\delta(1-x_{k})\Dlxi\\
&+\delta(1-x_{i})\Dlxii
+\left(\frac{1-x_{i}}{2}+\frac{\log(\frac{2}{1+x_{i}})}{1-x_{i}}-\frac{1+x_{i}}{2}\log(\frac{2(1-x_{i})}{1+x_{i}})\right)\delta(1-x_{k})\\
&+\left(\frac{\log(\frac{2}{1+x_{k}})}{1-x_{k}}-\frac{(x_{k}^{3}-x_{k}^{2}+2x_{k}-1)\log(\frac{2(1-x_{k})}{1+x_{k}})}{x_{k}}\right)\delta(1-x_{i})-\frac{1+x_{i}}{2}\Doxii\\
&+\left(-x_{k}^{2}+x_{k}-2+\frac{1}{x_{k}}\right)\Doxi+\Doxi\Doxii\\
&\frac{Y_2(x_{i},x_{k})-Y_2(x_{i},1)-Y_2(1,x_{k})+Y_2(1,1)}{(1-x_{i})(1-x_{k})}\Big)+\order{\e},
\end{split}
\end{equation}
where
\begin{equation}
Y_2(x_{i},x_{k})=\frac{(1+x_{i}x_{k})(x_{i}x_{k}^{3}(1-x_{i}^{2})^{2}+2x_{i}^{2}(1+x_{i}^2x_k^2)(1-x_{k}^{2})^{2}+2x_{i}^2x_{k}^{2}(x_{i}+x_{k})^{2})}
{(x_{i}+x_{k})^{2}(1+x_{i})(1+x_{k})x_{i}x_{k}}.
\end{equation}
This agrees with the result of Ref.~\cite{Daleo:2006xa} up to non-singular terms at $\order{\e^0}$.

\subsubsection{$\ggF(\widehat{i}_{g},j_{g},\widehat{k}_{g})$}
The $ggg$ antenna with two colour-unconnected gluons ($i$ and $k$) in the initial state has limits when the colour connected gluon ($j$) is soft, or collinear with either of the initial-state gluons. 
The antenna constructed from these limits is given by
\begin{equation}
\label{eq:ggF30II}
\ggF(\widehat{i}_{g},j_{g},\widehat{k}_{g})= 
\frac{2\sik}{\sij\sjk}
+\frac{2\sjk}{\sij\sik}
+\frac{2\sjk\sik}{\sij\sijk^{2}}
+\frac{2\sij}{\sik\sjk}
+\frac{2\sij\sik}{\sjk\sijk^{2}}.
\end{equation}
Just as with the $\qgD$ antenna, this antenna cannot be obtained by crossing the $\F$ antenna for final-final kinematics given in Ref.~\cite{paper2} because it includes the full $\widehat{g}g$ collinear limit, rather than the $g^hg$ collinear limit.
Comparing with the antenna given in Ref.~\cite{Daleo:2006xa},
\begin{equation}
\ggFold(\widehat{i}_{g},j_{g},\widehat{k}_{g})=\ggF(\widehat{i}_{g},j_{g},\widehat{k}_{g})+\frac{2\sjk\sij}{\sik\sijk^{2}}+\frac{8\sik}{\sijk^{2}}+\frac{8\sjk}{\sijk^{2}}+\frac{8\sij}{\sijk^{2}}+\order{\e}
\end{equation}
we see that they agree up to $\order{\e}$ and non-singular terms.

Integrating Eq.~\eqref{eq:ggF30II} over the initial-initial antenna phase space yields
\begin{equation}
\begin{split}
\calggF(\scaleII,\xi,\xk) &= -\bold{I}_{gg}^{(1)}(\e,\scaleII)\delta(1-x_{i})\delta(1-x_{k})+\QQII\Big(-\frac{1}{2\e}p_{gg}^{(0)}(x_{i})\delta(1-x_{k})\\
&+\delta(1-x_{i})\Dlxii+\frac{\pi^{2}}{4}\delta(1-x_{i})\delta(1-x_{k})\\
&-(x_{k}^{2}-x_{k}+2-\frac{1}{x_{k}})\log(\frac{2(1-x_{k})}{1+x_{k}})\delta(1-x_{i})+\frac{\log(\frac{2}{1+x_{k}})}{1-x_{k}}\delta(1-x_{i})\\
& - \left(x_{k}^{2}-x_{k}+2-\frac{1}{x_{k}}\right)\Doxi+\frac{1}{2}\Doxi\Doxii\\
&+\frac{Y_3(x_{i},x_{k})-Y_3(x_{i},1)-Y_3(1,x_{k})+Y_3(1,1)}{2(1-x_{i})(1-x_{k})}\Big)+(x_{i}\leftrightarrow x_{k})+\order{\e},
\end{split}
\end{equation}
where
\begin{equation}
Y_3(x_{i},x_{k})=\frac{(1+x_{i}x_{k})(x_{i}^{2}x_{k}^{2}(x_{i}+x_{k})^{2}
+2x_i^2(1+x_{i}^2x_{k}^2)(1-x_{k}^{2})^{2})}{x_{i}x_{k}(1+x_{i})(1+x_{k})(x_{i}+x_{k})^{2}}+(x_{i}\leftrightarrow x_{k}).
\end{equation}
This agrees with the result of Ref.~\cite{Daleo:2006xa} up to non-singular terms at $\order{\e^0}$.

\subsection{Identity-changing initial-initial antennae}

As indicated in Table~\ref{tab:X30II}, there are four identity-changing initial-initial antenna: two describing quark to gluon transitions and two describing quark to gluon transitions.

\subsubsection{$\qgA(\widehat{i}_{q},\widehat{k}_{g},j_{\bar{q}})$}

The identity-changing $qg\bar{q}$ antenna, $\qgA(\widehat{i}_{q},\widehat{k}_{g},j_{\bar{q}})$, only has a limit when the final-state  quark is collinear with the initial-state gluon.  
The algorithm yields
\begin{equation}
\label{eq:qgA30II}
\qgA(\widehat{i}_{q},\widehat{k}_{g},j_{\bar{q}})= -\frac{\sik\ome}{\sijk\sjk}-\frac{2\sij}{\sik\sjk}.
\end{equation}
Comparing to the antenna of Ref.~\cite{Daleo:2006xa}
\begin{equation}
\qgAold(\widehat{i}_{q},\widehat{k}_{g},j_{\bar{q}})= \qgA(\widehat{i}_{q},\widehat{k}_{g},j_{\bar{q}}) -\frac{\sjk}{\sijk\sik} + \order{\e},
\end{equation}
we see the difference is either higher order in $\e$ or terms that are non-singular in phase space like $\frac{1}{\sik}$. In this configuration, $\sik$ cannot go to zero as both particles $i$ and $k$ are in the initial state.

Integrating Eq.~\eqref{eq:qgA30II} over the initial-initial antenna phase space yields
\begin{equation}
\begin{split}
&\calqgA(\scaleII,\xi,\xk) = \QQII\Big(-\frac{1}{2\e}p_{qg}^{(0)}(x_{k})\delta(1-x_{i})+\frac{2x_{k}^{2}-2x_{k}+1}{2}\Doxi\\
&+\left(\frac{1}{2}+\frac{2x_{k}^{2}-2x_{k}+1}{2}\log\left(\frac{2(1-x_{k})}{1+x_{k}}\right)\right)\delta(1-x_{i})+\frac{Y_4(x_{i},x_{k})-Y_4(1,x_{k})}{1-x_{i}}\Big)+\order{\e},
\end{split}
\end{equation}
where
\begin{equation}
\label{eq:Y4}
Y_4(x_{i},x_{k})=\frac{x_{i}(1+x_{i}x_{k})(-2x_{i}^{2}x_{k}(1-x_{k}^{2})+(x_{i}+x_{k}))}{(x_{i}+x_{k})^{2}(1+x_{i})}.
\end{equation}
This agrees with the result of Ref.~\cite{Daleo:2006xa} up to non-singular terms at $\order{\e^0}$.

\subsubsection{$\ggD(j_q,\widehat{k}_{g},\widehat{i}_{g})$}
The identity-changing $\ggD(j_q,\widehat{k}_{g},\widehat{i}_{g})$ antenna only has a limit when the final-state  quark is collinear with the adjacent initial-state gluon $k$.
This leads to
\begin{equation}
\label{eq:ggD30II}
\ggD(j_q,\widehat{k}_{g},\widehat{i}_{g}) = 
-\frac{\sik\ome}{\sijk\sjk}
-\frac{2\sij}{\sik\sjk}.
\end{equation}
This is precisely the same as Eq.~\eqref{eq:qgA30II}, because it has the same collinear limit in which the initial-state particle transitions from a gluon to a quark.
The corresponding antenna in Ref.~\cite{Daleo:2006xa} includes the collinear limits of the quark with either initial-state gluon, so that
\begin{equation}
\begin{split}
\ggDold(j_q,\widehat{i}_{g},\widehat{k}_{g})&=        \ggD(j_q,\widehat{k}_{g},\widehat{i}_{g})
+ \ggD(j_q,\widehat{i}_{g},\widehat{k}_{g}) \\
&-\frac{2\sjk\sij}{\sik\sijk^{2}}-\frac{5\sjk}{\sijk^{2}}-\frac{5\sij}{\sijk^{2}}-\frac{4\sik}{\sijk^{2}} +\order{\e}.
\end{split}
\end{equation}

Integrating Eq.~\eqref{eq:ggD30II} over the initial-initial antenna phase space yields
\begin{equation}
\begin{split}
&\calggD(\scaleII,\xi,\xk) = \QQII\Big(-\frac{1}{2\e}p_{qg}^{(0)}(x_{k})\delta(1-x_{i})+\frac{2x_{k}^{2}-2x_{k}+1}{2}\Doxi\\
&+\left(\frac{1}{2}+\frac{2x_{k}^{2}-2x_{k}+1}{2}\log\left(\frac{2(1-x_{k})}{1+x_{k}}\right)\right)\delta(1-x_{i})+\frac{Y_4(x_{i},x_{k})-Y_4(1,x_{k})}{1-x_{i}}\Big)+\order{\e},
\end{split}
\end{equation}
with $Y_4$ given in Eq.~\eqref{eq:Y4}.
This agrees with the result of Ref.~\cite{Daleo:2006xa} up to non-singular terms at $\order{\e^0}$.

\subsubsection{$\qQE(\widehat{i}_{q},\widehat{k}_{\qpbar},j_{\qp})$}
The identity-changing $\qQE(\widehat{i}_{q},\widehat{k}_{\qpbar},j_{\qp})$ antenna only has a limit when the final-state  quark is collinear with the initial-state quark of the same flavour, and is described by the $P_{gq\leftarrow Q}$ splitting function.
Applying the algorithm yields
\begin{equation}
\label{eq:qQE30II}
\qQE(\widehat{i}_{q},\widehat{k}_{\qpbar},j_{\qp})= -\frac{1}{\sjk}+\frac{2\sik\sij}{\sjk\sijk^{2}\ome}.
\end{equation}
This is identical to the $\gQG$ antenna, since it describes the same limit.
It agrees with the antenna of Ref.~\cite{Daleo:2006xa} up to $\order{\e}$ and non-singular terms,
\begin{equation}
\qQEold(\widehat{i}_{g},\widehat{k}_{\qpbar},j_{\qp})= \qQE(\widehat{i}_{g},\widehat{k}_{\qpbar},j_{\qp})+\frac{\sij}{\sijk^{2}}+\frac{\sjk}{\sijk^{2}}+\frac{\sik}{\sijk^{2}}+\order{\e}.
\end{equation}
Eq.~\eqref{eq:qQE30II} can also be obtained by crossing Eq.~\eqref{eq:QtogE30IF} with the exchange $i \leftarrow k$.

Integrating Eq.~\eqref{eq:qQE30II} over the initial-initial antenna phase space yields
\begin{equation}
\begin{split}
\calqQE(\scaleII,\xi,\xk)&=\QQII\Big(-\frac{1}{2\e}p_{gq}^{(0)}(x_{k})\delta(1-x_{i})\\
&+\frac{2x_{k}-2 +(2-2x_{k}+x_{k}^{2})\text{log}(\frac{2(1-x_{k})}{1+x_{k}})}{2x_{k}}\delta(1-x_{i})\\
&+\frac{x_{k}^{2}-2x_{k}+2}{2x_{k}}\Doxi+\frac{Y_5(x_{i},x_{k})-Y_5(1,x_{k})}{(1-x_{i})}\Big)+\order{\e},
\end{split}
\end{equation}
where
\begin{equation}
\label{eq:Y5}
Y_5(x_{i},x_{k})=\frac{(1+x_{i}x_{k})(x_{i}^{2}x_{k}^{2}(x_{i}+x_{k})+2x_{i}(1-x_{k}^{2}))}{x_{k}(x_{i}+x_{k})^{2}(1+x_{i})}.
\end{equation}
This agrees with the result of Ref.~\cite{Daleo:2006xa} up to non-singular terms at $\order{\e^0}$. 

\subsubsection{$\gQG(\widehat{i}_{g},\widehat{k}_{\qpbar},j_{\qp})$}
The identity-changing $\gQG(\widehat{i}_{g},\widehat{k}_{\qpbar},j_{\qp})$ antenna only has a limit when the final-state  quark is collinear with the initial-state quark of the same flavour.
Applying the algorithm yields
\begin{equation}
\label{eq:gQG30II}
\gQG(\widehat{i}_{g},\widehat{k}_{\qpbar},j_{\qp})= -\frac{1}{\sjk}+\frac{2\sik\sij}{\sjk\sijk^{2}\ome}.
\end{equation}
This is identical to the $\qQE$ antenna, since it describes the same limit and agrees with the antenna of Ref.~\cite{Daleo:2006xa} up to $\order{\e}$ and non-singular terms,
\begin{equation}
\gQGold(\widehat{i}_{g},\widehat{k}_{\qpbar},j_{\qp})= \gQG(\widehat{i}_{g},\widehat{k}_{\qpbar},j_{\qp})+\frac{\sjk}{\sijk^{2}}+\frac{2\sij}{\sijk^{2}}+\frac{2\sik}{\sijk^{2}}+\order{\e}.
\end{equation}
Eq.~\eqref{eq:gQG30II} can also be obtained by crossing Eq.~\eqref{eq:QtogG30IF} with the exchange $i \leftarrow k$.

Integrating Eq.~\eqref{eq:gQG30II} over the initial-initial antenna phase space yields
\begin{equation}
\begin{split}
\calgQG(\scaleII,\xi,\xk)&=\QQII\Big(-\frac{1}{2\e}p_{gq}^{(0)}(x_{k})\delta(1-x_{i})\\
&+\frac{2x_{k}-2 +(2-2x_{k}+x_{k}^{2})\text{log}(\frac{2(1-x_{k})}{1+x_{k}})}{2x_{k}}\delta(1-x_{i})\\
&+\frac{x_{k}^{2}-2x_{k}+2}{2x_{k}}\Doxi+\frac{Y_5(x_{i},x_{k})-Y_5(1,x_{k})}{(1-x_{i})}\Big)+\order{\e},
\end{split}
\end{equation}
with $Y_5$ given in Eq.~\eqref{eq:Y5}.
This agrees with the result of Ref.~\cite{Daleo:2006xa} up to non-singular terms at $\order{\e^0}$.

\section{Validation}
\label{sec:validation}

\begin{figure}[th]
    \centering
    \includegraphics[width=0.45\textwidth]{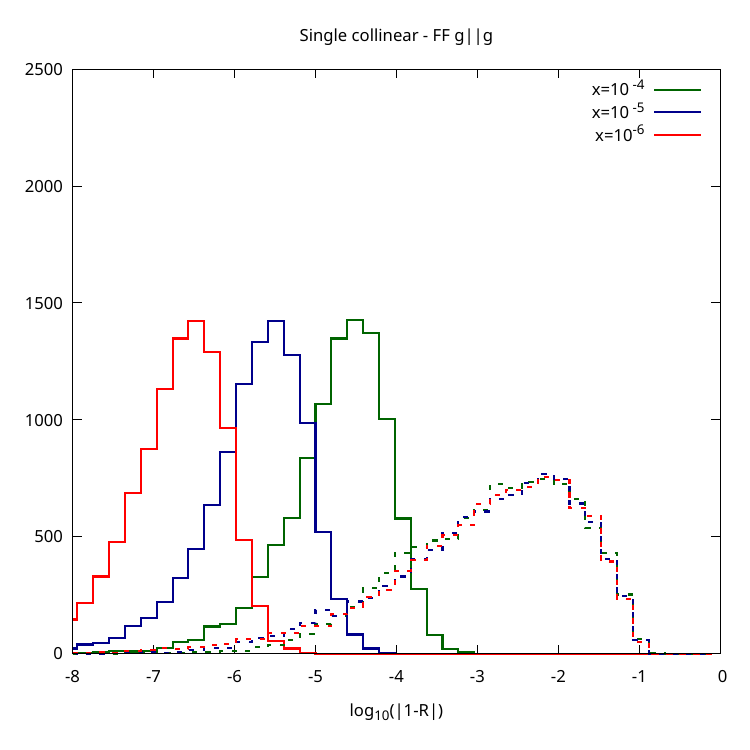}
    \includegraphics[width=0.45\textwidth]{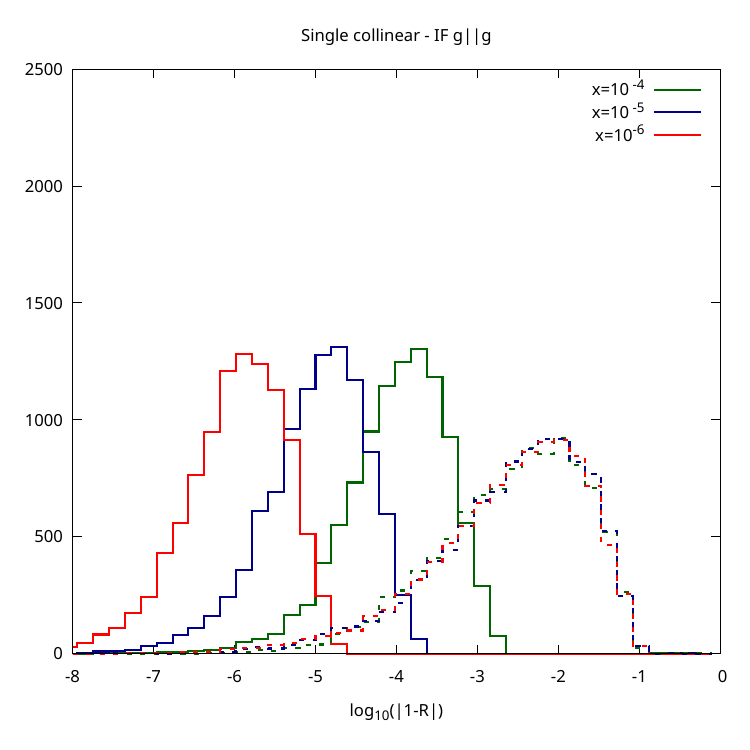}
    \includegraphics[width=0.45\textwidth]{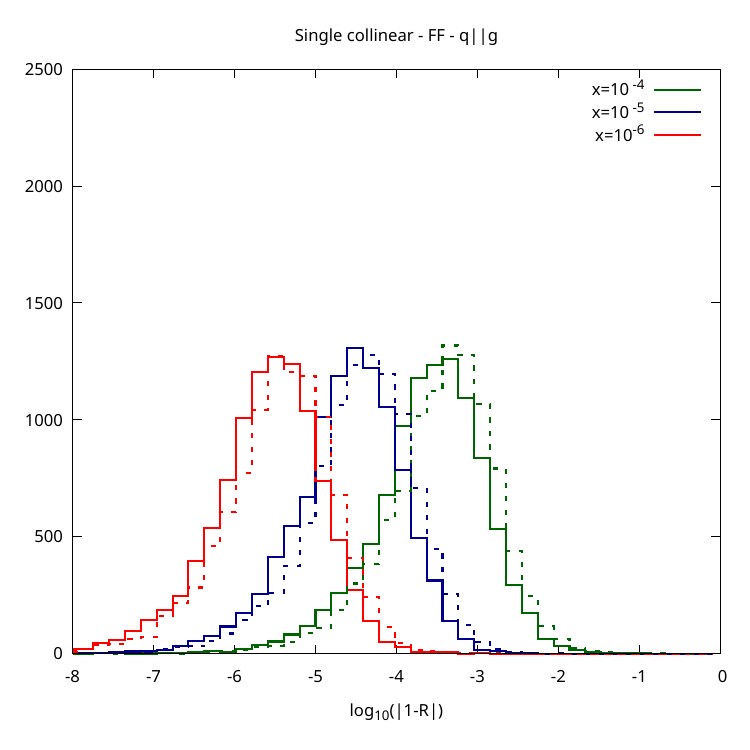}   
    \includegraphics[width=0.45\textwidth]{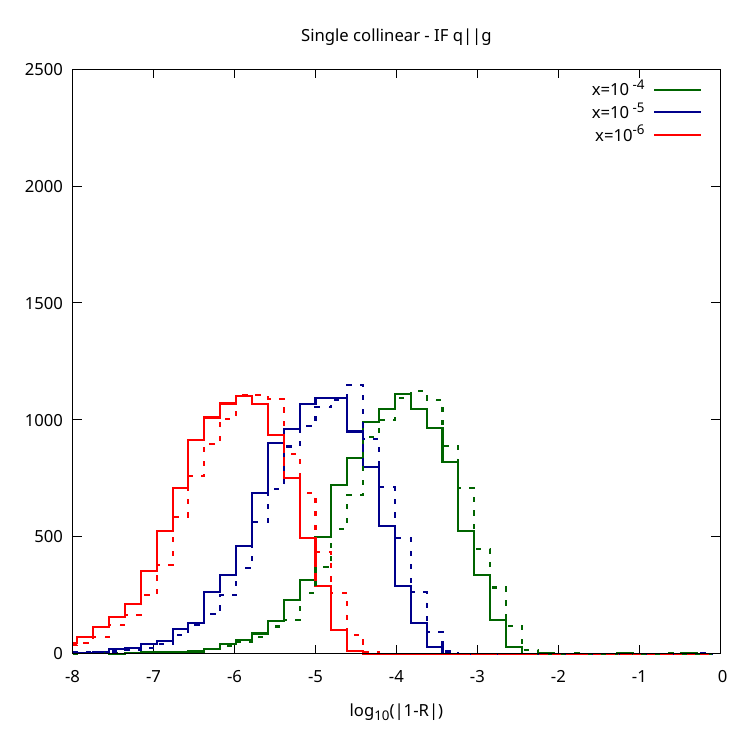}   
    \caption{The deviation of the ratio $R$ from unity is shown for two collinear gluons (upper frames) and a collinear quark-gluon pair (lower frames) in the final-final configuration (left frames) and initial-final configuration (right frames) for three values of the scaling parameter $x$. The solid (dashed) lines show the result combining (not-combining) phase space points related by an azimuthal rotation of $\pi/2$.}
    \label{fig:azimuthalspiketests}
\end{figure}

\NG{In this section we revisit our earlier assertion that the collinear limits of matrix elements can be correctly described by the antenna functions based on spin-averaged splitting functions.  We focus on the NLO
subtraction terms $S$ for the $gg \to gggg$ and $qg \to gggq$ processes when only three jets are visible in the final state. We consider the ratio $R = |M|^2/S$ and examine the deviation of $R$ from unity in the unresolved region. As we approach any unresolved region, $R$ should approach 1.
Fig.~\ref{fig:azimuthalspiketests} shows the logarithmic distribution of the absolute value of $1-R$ for both final-final and initial-final kinematics.  In each case, we show the result when two gluons are collinear (and where azimuthal effects may be expected) and when a quark and gluon are collinear (in which case we do not expect any azimuthal effects). 
The approach to the collinear limit is controlled by a variable $x$, such that the collinear invariant $|s_{ij}| = x s_{12}$.  The smaller the value of $x$, the closer one approaches the collinear limit. In all plots, the dashed lines shows the result obtained using the phase space point, while the solid lines show the effect of combining the original phase space point with one rotated by $\pi/2$ about the collinear direction.
}

\NG{Let us first focus on the lower plots.  In both FF and IF configurations we observe that the $qg$ collinear limit is described very well.   The effect of azimuthal averaging is completely negligible.   This is exactly as expected.  The azimuthal terms are produced when a spin-one gluon splits into either a $q\bar{q}$ or $gg$ pair.   There are no azimuthal terms when a $q$ splits into a $qg$ pair. Furthermore, we see that as $x$ gets smaller, $R$ gets closer to unity.  The peak in the distribution moves left by roughly one unit for each factor of 1/10 in $x$.
}

\NG{On the other hand, the upper plots show something rather different.   Focussing first on the dashed lines, which represent the subtraction term at a given phase space point, we see that $R$ does not approach unity for smaller values of $x$.  In fact the subtraction term is a poor representation of the matrix element for all $x$ values. This is the case for both FF and IF configurations.
}

\NG{On the other hand, once the azimuthally related points are included, the subtraction term does successfully reproduce the matrix elements.  We see that the smaller the value of $x$, the better the subtraction term agrees with the matrix element.   Again, this is the case for both FF and IF configurations.
In fact, with this trick of combining azimuthal pairs, we see that the $gg$ collinear limit is better in the sense that the subtraction term is closer to the matrix elements for the $gg$ collinear limit than for the $qg$ collinear limit. }

\section{Outlook}
\label{sec:outlook}
We have extended the algorithm for constructing real-radiation antenna functions directly from their desired unresolved limits, as described in Ref.~\cite{paper2}, to include scenarios where one or both of the hard radiators are in the initial state. With this advancement, we derived a comprehensive set of \NG{single unresolved} initial-final and initial-initial antennae \NG{for massless partons}. We provide expressions for all the antennae and their integrated forms over the relevant antenna phase space. \NG{We expect that the generalisation to massive partons is straightforward. We demonstrated numerically that the antennae based on spin-averaged collinear splitting functions do describe the collinear limits of single unresolved multi-particle matrix elements, once pairs of events rotated by $\pi/2$ about the collinear direction are combined.  Our work} marks a significant step towards a more streamlined antenna subtraction scheme \NG {at NNLO and beyond}, capable of calculating higher-order QCD corrections to exclusive collider observables involving partons in the initial state.

Unlike the case when all hard radiators are in the final state, ensuring that the algorithm only generates denominators corresponding to physical propagators becomes challenging for initial-final antennae. Careful attention is required to avoid introducing singularities in the antennae when the initial-state and final-state hard radiators become collinear. This leads to the appearance of composite denominators, but it does facilitate a clear separation between flavor-preserving and flavor-changing antennae. Nonetheless, the \NG{single unresolved} antennae presented here can be integrated analytically, which is a key feature of the antenna-subtraction method. In all instances, we observe consistent agreement with the \NG{single unresolved} antennae presented in Ref.~\cite{Daleo:2006xa}, up to terms that remain finite in the unresolved regions for the unintegrated antennae, and up to terms of $\order{\e^0}$ for the integrated antennae.

The focus of the algorithm on singular limits leads to a reduction in the number of \NG{single unresolved} antennae needed at NLO. 
There are 22 possible antennae listed in Tables~\ref{tab:X30FF}, \ref{tab:X30IF} and \ref{tab:X30II}.  For the fully differential (unintegrated) antennae, this can be reduced to ten independent functions - four Final-Final ($\A,~\D,~\E,~\F$), three Initial-Final (two identity-preserving $\gD,~\gF$ and one identity-changing $\gtoqA$) and three Initial-Initial (two identity-preserving $\qgD,~\ggF$ and one identity-changing $\qgA$).  As listed in Table~\ref{tab:X30independent}, the other 12 are either equivalent or can be obtained by crossing. 

Once the integration over the antenna phase space is performed, then we need sixteen functions: four Final-Final ($\calA,~\calD,~\calE,~\calF$), seven Initial-Final (five identity-preserving $\calqA$, $\calqD$, $\calqE$, $\calgD$,~$\calgF$ and two identity-changing $\calgtoqA,~\calQtogE$) and five Initial-Initial (three identity-preserving $\calqqA, \calqgD,~\calggF$ and two identity-changing $\calqgA$ and $\calqQE$).

\NG{Since a large part of the NNLO subtraction term consists of products of $X_3^0$ and/or ${\cal X}_3^0$ antennae, we expect that the size of the subtraction term constructed using the new $X_3^0$ and ${\cal X}_3^0$ antenna will be reduced.}

\begin{table}[t]
\centering
\begin{tabular}{ll}
\bf{Final-Final Antennae}  & \\
$\G$ & $\equiv \E$\\
\bf{Initial-Final Antennae}  & \\
$\qA$  & crossing of $\A$\\
$\qD$  & crossing of $\D$\\
$\qE$  & crossing of $\E$\\
$\gG$  & $\equiv \qE$\\
$\gtoqD$  & $\equiv \gtoqA$\\
$\QtogE$  & crossing of $\E$\\
$\QtogG$  & $\equiv \QtogE$\\
\bf{Initial-Initial Antennae}  & \\
$\qqA$  & crossing of $\A$\\
$\ggD$   & $\equiv \qgA$\\
$\qQE$   & crossing of $\QtogE$ \\
$\gQG$   & $\equiv \qQE$
\end{tabular}
\caption{The relationship between the full set of three particle antennae and the ten independent antennae. }
\label{tab:X30independent}
\end{table}

We anticipate that the technique presented here for building antenna functions with radiators in the initial state directly from the required limits, along with the constructed antenna functions, will not only simplify the antenna-subtraction framework significantly, paving the way for full automation, but will also find applications in parton showers and their matching to NNLO calculations. 
The ability to directly integrate the antennae contributes to the versatility and efficiency of the approach, making it a valuable tool for future studies in precision collider phenomenology.

\acknowledgments
We thank Oscar Braun-White, Xuan Chen, Aude Gehrmann-De Ridder, Thomas Gehrmann, Matteo Marcoli and Christian Preuss for enlightening discussions and helpful advice. We thank the University of Zurich, and especially Thomas Gehrmann and his research group for their kind hospitality, while visiting the University of Zurich. This visit was supported in part by the Pauli Center for Theoretical Studies, in part by the UK Science and Technology Facilities Council under contract ST/T001011/1 and in part by the Swiss National Science Foundation under contract 200021-197130.
\appendix

\section{Colour-Ordered Infrared Singularity Operators}
The colour-ordered infrared singularity operators are as in
\cite{Catani:1998nv}.
\label{sec:catani}
\begin{equation}
\begin{split}
\bold{I}^{(1)}_{q\bar{q}}(\e,s)&=-\frac{e^{\e\gamma}}{2\Gamma(1-\e)}\Big(\frac{1}{\e^{2}}+\frac{3}{2\e}\Big)\mathcal{R}((-s)^{-\e}),\\
\bold{I}^{(1)}_{qg}(\e,s)&=-\frac{e^{\e\gamma}}{2\Gamma(1-\e)}\Big(\frac{1}{\e^{2}}+\frac{5}{3\e}\Big)\mathcal{R}((-s)^{-\e}),\\
\bold{I}^{(1)}_{gg}(\e,s)&=-\frac{e^{\e\gamma}}{2\Gamma(1-\e)}\Big(\frac{1}{\e^{2}}+\frac{11}{6\e}\Big)\mathcal{R}((-s)^{-\e}),\\
\bold{I}^{(1)}_{q\bar{q},F}(\e,s)&=0,\\
\bold{I}^{(1)}_{qg,F}(\e,s)&=\frac{e^{\e\gamma}}{2\Gamma(1-\e)}\frac{1}{6\e}\mathcal{R}((-s)^{-\e}),\\
\bold{I}^{(1)}_{gg,F}(\e,s)&=\frac{e^{\e\gamma}}{2\Gamma(1-\e)}\frac{1}{3\e}\mathcal{R}((-s)^{-\e}).\\
\end{split}
\end{equation}

\section{Colour-ordered splitting kernels}
\label{sec:kernels}

The colour ordered splitting kernels are given by \cite{Daleo:2006xa},
\begin{equation}
\begin{split}
p_{qq}^{(0)}(x)&=\frac{3}{2}\delta(1-x)+2\mathcal{D}_{0}(x)-1-x,\\
p_{qg}^{(0)}(x)&=1-2x+2x^{2},\\
p_{gq}^{(0)}(x)&=\frac{2}{x}-2+x,\\
p_{gg}^{(0)}(x)&=\frac{11}{6}\delta(1-x)+2\mathcal{D}_{0}(x)+\frac{2}{x}-4+2x-2x^{2},\\
p_{gg,F}^{(0)}(x)&=-\frac{1}{3}\delta(1-x).
\end{split}
\end{equation}

\bibliographystyle{jhep}
\bibliography{bib2}{}

\providecommand{\href}[2]{#2}\begingroup\raggedright\begin{thebibliography}{100}

\bibitem{paper2}
O.~Braun-White, N.~Glover, and C.~T. Preuss, {\it {A general algorithm to build
  real-radiation antenna functions for higher-order calculations}},  {\em JHEP}
  {\bf 06} (2023) 065, [\href{http://arxiv.org/abs/2302.12787}{{\tt
  arXiv:2302.12787}}].

\bibitem{Kinoshita}
T.~Kinoshita, {\it Mass singularities of feynman amplitudes},  {\em Journal of
  Mathematical Physics} {\bf 3} (1962), no.~4 650--677,
  [\href{http://arxiv.org/abs/https://doi.org/10.1063/1.1724268}{{\tt
  https://doi.org/10.1063/1.1724268}}].

\bibitem{LeeNauenberg}
T.~D. Lee and M.~Nauenberg, {\it Degenerate systems and mass singularities},
  {\em Phys. Rev.} {\bf 133} (Mar, 1964) B1549--B1562.

\bibitem{Catani:1996vz}
S.~Catani and M.~H. Seymour, {\it {A General algorithm for calculating jet
  cross-sections in NLO QCD}},  {\em Nucl. Phys. B} {\bf 485} (1997) 291--419,
  [\href{http://arxiv.org/abs/hep-ph/9605323}{{\tt hep-ph/9605323}}]. [Erratum:
  Nucl.Phys.B 510, 503--504 (1998)].

\bibitem{Catani:1996jh}
S.~Catani and M.~H. Seymour, {\it {The Dipole formalism for the calculation of
  QCD jet cross-sections at next-to-leading order}},  {\em Phys. Lett. B} {\bf
  378} (1996) 287--301, [\href{http://arxiv.org/abs/hep-ph/9602277}{{\tt
  hep-ph/9602277}}].

\bibitem{Catani:2002hc}
S.~Catani, S.~Dittmaier, M.~H. Seymour, and Z.~Trocsanyi, {\it {The Dipole
  formalism for next-to-leading order QCD calculations with massive partons}},
  {\em Nucl. Phys. B} {\bf 627} (2002) 189--265,
  [\href{http://arxiv.org/abs/hep-ph/0201036}{{\tt hep-ph/0201036}}].

\bibitem{Frixione:1995ms}
S.~Frixione, Z.~Kunszt, and A.~Signer, {\it {Three jet cross-sections to
  next-to-leading order}},  {\em Nucl. Phys. B} {\bf 467} (1996) 399--442,
  [\href{http://arxiv.org/abs/hep-ph/9512328}{{\tt hep-ph/9512328}}].

\bibitem{Frixione:1997np}
S.~Frixione, {\it {A General approach to jet cross-sections in QCD}},  {\em
  Nucl. Phys. B} {\bf 507} (1997) 295--314,
  [\href{http://arxiv.org/abs/hep-ph/9706545}{{\tt hep-ph/9706545}}].

\bibitem{Frederix:2008hu}
R.~Frederix, T.~Gehrmann, and N.~Greiner, {\it {Automation of the Dipole
  Subtraction Method in MadGraph/MadEvent}},  {\em JHEP} {\bf 09} (2008) 122,
  [\href{http://arxiv.org/abs/0808.2128}{{\tt arXiv:0808.2128}}].

\bibitem{Frederix:2009yq}
R.~Frederix, S.~Frixione, F.~Maltoni, and T.~Stelzer, {\it {Automation of
  next-to-leading order computations in QCD: The FKS subtraction}},  {\em JHEP}
  {\bf 10} (2009) 003, [\href{http://arxiv.org/abs/0908.4272}{{\tt
  arXiv:0908.4272}}].

\bibitem{Frederix:2010cj}
R.~Frederix, T.~Gehrmann, and N.~Greiner, {\it {Integrated dipoles with
  MadDipole in the MadGraph framework}},  {\em JHEP} {\bf 06} (2010) 086,
  [\href{http://arxiv.org/abs/1004.2905}{{\tt arXiv:1004.2905}}].

\bibitem{madgraph:2011uj}
J.~Alwall, M.~Herquet, F.~Maltoni, O.~Mattelaer, and T.~Stelzer, {\it {MadGraph
  5 : Going Beyond}},  {\em JHEP} {\bf 06} (2011) 128,
  [\href{http://arxiv.org/abs/1106.0522}{{\tt arXiv:1106.0522}}].

\bibitem{Cascioli:2011va}
F.~Cascioli, P.~Maierhofer, and S.~Pozzorini, {\it {Scattering Amplitudes with
  Open Loops}},  {\em Phys. Rev. Lett.} {\bf 108} (2012) 111601,
  [\href{http://arxiv.org/abs/1111.5206}{{\tt arXiv:1111.5206}}].

\bibitem{Frixione:2002ik}
S.~Frixione and B.~R. Webber, {\it {Matching NLO QCD computations and parton
  shower simulations}},  {\em JHEP} {\bf 06} (2002) 029,
  [\href{http://arxiv.org/abs/hep-ph/0204244}{{\tt hep-ph/0204244}}].

\bibitem{Nason:2004rx}
P.~Nason, {\it {A New method for combining NLO QCD with shower Monte Carlo
  algorithms}},  {\em JHEP} {\bf 11} (2004) 040,
  [\href{http://arxiv.org/abs/hep-ph/0409146}{{\tt hep-ph/0409146}}].

\bibitem{Frixione:2007vw}
S.~Frixione, P.~Nason, and C.~Oleari, {\it {Matching NLO QCD computations with
  Parton Shower simulations: the POWHEG method}},  {\em JHEP} {\bf 11} (2007)
  070, [\href{http://arxiv.org/abs/0709.2092}{{\tt arXiv:0709.2092}}].

\bibitem{powheg:2010xd}
S.~Alioli, P.~Nason, C.~Oleari, and E.~Re, {\it {A general framework for
  implementing NLO calculations in shower Monte Carlo programs: the POWHEG
  BOX}},  {\em JHEP} {\bf 06} (2010) 043,
  [\href{http://arxiv.org/abs/1002.2581}{{\tt arXiv:1002.2581}}].

\bibitem{Bellm:2019zci}
J.~Bellm et~al., {\it {Herwig 7.2 release note}},  {\em Eur. Phys. J. C} {\bf
  80} (2020), no.~5 452, [\href{http://arxiv.org/abs/1912.06509}{{\tt
  arXiv:1912.06509}}].

\bibitem{Sherpa:2019gpd}
{\bf Sherpa} Collaboration, E.~Bothmann et~al., {\it {Event Generation with
  Sherpa 2.2}},  {\em SciPost Phys.} {\bf 7} (2019), no.~3 034,
  [\href{http://arxiv.org/abs/1905.09127}{{\tt arXiv:1905.09127}}].

\bibitem{Bierlich:2022pfr}
C.~Bierlich et~al., {\it {A comprehensive guide to the physics and usage of
  PYTHIA 8.3}},  \href{http://arxiv.org/abs/2203.11601}{{\tt
  arXiv:2203.11601}}.

\bibitem{Nagy:2003qn}
Z.~Nagy and D.~E. Soper, {\it {General subtraction method for numerical
  calculation of one loop QCD matrix elements}},  {\em JHEP} {\bf 09} (2003)
  055, [\href{http://arxiv.org/abs/hep-ph/0308127}{{\tt hep-ph/0308127}}].

\bibitem{Bevilacqua:2013iha}
G.~Bevilacqua, M.~Czakon, M.~Kubocz, and M.~Worek, {\it {Complete Nagy-Soper
  subtraction for next-to-leading order calculations in QCD}},  {\em JHEP} {\bf
  10} (2013) 204, [\href{http://arxiv.org/abs/1308.5605}{{\tt
  arXiv:1308.5605}}].

\bibitem{Prisco:2020kyb}
R.~M. Prisco and F.~Tramontano, {\it {Dual subtractions}},  {\em JHEP} {\bf 06}
  (2021) 089, [\href{http://arxiv.org/abs/2012.05012}{{\tt arXiv:2012.05012}}].

\bibitem{Bertolotti:2022ohq}
G.~Bertolotti, P.~Torrielli, S.~Uccirati, and M.~Zaro, {\it {Local analytic
  sector subtraction for initial- and final-state radiation at NLO in massless
  QCD}},  {\em JHEP} {\bf 12} (2022) 042,
  [\href{http://arxiv.org/abs/2209.09123}{{\tt arXiv:2209.09123}}].

\bibitem{Giachino:2023loc}
A.~Giachino, A.~van Hameren, and G.~Ziarko, {\it {A new subtraction scheme at
  NLO exploiting the privilege of kT-factorization}},
  \href{http://arxiv.org/abs/2312.02808}{{\tt arXiv:2312.02808}}.

\bibitem{Anastasiou:2003gr}
C.~Anastasiou, K.~Melnikov, and F.~Petriello, {\it {A new method for real
  radiation at NNLO}},  {\em Phys. Rev. D} {\bf 69} (2004) 076010,
  [\href{http://arxiv.org/abs/hep-ph/0311311}{{\tt hep-ph/0311311}}].

\bibitem{Frixione:2004is}
S.~Frixione and M.~Grazzini, {\it {Subtraction at NNLO}},  {\em JHEP} {\bf 06}
  (2005) 010, [\href{http://arxiv.org/abs/hep-ph/0411399}{{\tt
  hep-ph/0411399}}].

\bibitem{Gehrmann-DeRidder:2005btv}
A.~Gehrmann-De~Ridder, T.~Gehrmann, and E.~W.~N. Glover, {\it {Antenna
  subtraction at {NNLO}}},  {\em JHEP} {\bf 09} (2005) 056,
  [\href{http://arxiv.org/abs/hep-ph/0505111}{{\tt hep-ph/0505111}}].

\bibitem{Gehrmann-DeRidder:2007foh}
A.~Gehrmann-De~Ridder, T.~Gehrmann, E.~W.~N. Glover, and G.~Heinrich, {\it
  {Infrared structure of $e^+e^- \to 3$ jets at NNLO}},  {\em JHEP} {\bf 11}
  (2007) 058, [\href{http://arxiv.org/abs/0710.0346}{{\tt arXiv:0710.0346}}].

\bibitem{Currie:2013vh}
J.~Currie, E.~W.~N. Glover, and S.~Wells, {\it {Infrared Structure at NNLO
  Using Antenna Subtraction}},  {\em JHEP} {\bf 04} (2013) 066,
  [\href{http://arxiv.org/abs/1301.4693}{{\tt arXiv:1301.4693}}].

\bibitem{Catani:2007vq}
S.~Catani and M.~Grazzini, {\it {An NNLO subtraction formalism in hadron
  collisions and its application to Higgs boson production at the LHC}},  {\em
  Phys. Rev. Lett.} {\bf 98} (2007) 222002,
  [\href{http://arxiv.org/abs/hep-ph/0703012}{{\tt hep-ph/0703012}}].

\bibitem{Czakon:2010td}
M.~Czakon, {\it {A novel subtraction scheme for double-real radiation at
  NNLO}},  {\em Phys. Lett. B} {\bf 693} (2010) 259--268,
  [\href{http://arxiv.org/abs/1005.0274}{{\tt arXiv:1005.0274}}].

\bibitem{Czakon:2014oma}
M.~Czakon and D.~Heymes, {\it {Four-dimensional formulation of the
  sector-improved residue subtraction scheme}},  {\em Nucl. Phys. B} {\bf 890}
  (2014) 152--227, [\href{http://arxiv.org/abs/1408.2500}{{\tt
  arXiv:1408.2500}}].

\bibitem{Caola:2017dug}
F.~Caola, K.~Melnikov, and R.~R\"ontsch, {\it {Nested soft-collinear
  subtractions in NNLO QCD computations}},  {\em Eur. Phys. J. C} {\bf 77}
  (2017), no.~4 248, [\href{http://arxiv.org/abs/1702.01352}{{\tt
  arXiv:1702.01352}}].

\bibitem{Somogyi:2006da}
G.~Somogyi, Z.~Trocsanyi, and V.~Del~Duca, {\it {A Subtraction scheme for
  computing QCD jet cross sections at NNLO: Regularization of doubly-real
  emissions}},  {\em JHEP} {\bf 01} (2007) 070,
  [\href{http://arxiv.org/abs/hep-ph/0609042}{{\tt hep-ph/0609042}}].

\bibitem{Somogyi:2006db}
G.~Somogyi and Z.~Trocsanyi, {\it {A Subtraction scheme for computing QCD jet
  cross sections at NNLO: Regularization of real-virtual emission}},  {\em
  JHEP} {\bf 01} (2007) 052, [\href{http://arxiv.org/abs/hep-ph/0609043}{{\tt
  hep-ph/0609043}}].

\bibitem{DelDuca:2016ily}
V.~Del~Duca, C.~Duhr, A.~Kardos, G.~Somogyi, Z.~Szor, Z.~{Tr\'ocs\'anyi}, and
  Z.~{Tulip\'ant}, {\it {Jet production in the CoLoRFulNNLO method: event
  shapes in electron-positron collisions}},  {\em Phys. Rev. D} {\bf 94}
  (2016), no.~7 074019, [\href{http://arxiv.org/abs/1606.03453}{{\tt
  arXiv:1606.03453}}].

\bibitem{Magnea:2018hab}
L.~Magnea, E.~Maina, G.~Pelliccioli, C.~Signorile-Signorile, P.~Torrielli, and
  S.~Uccirati, {\it {Local analytic sector subtraction at NNLO}},  {\em JHEP}
  {\bf 12} (2018) 107, [\href{http://arxiv.org/abs/1806.09570}{{\tt
  arXiv:1806.09570}}]. [Erratum: JHEP 06, 013 (2019)].

\bibitem{Herzog:2018ily}
F.~Herzog, {\it {Geometric IR subtraction for final state real radiation}},
  {\em JHEP} {\bf 08} (2018) 006, [\href{http://arxiv.org/abs/1804.07949}{{\tt
  arXiv:1804.07949}}].

\bibitem{TorresBobadilla:2020ekr}
W.~J. Torres~Bobadilla et~al., {\it {May the four be with you: Novel
  IR-subtraction methods to tackle NNLO calculations}},  {\em Eur. Phys. J. C}
  {\bf 81} (2021), no.~3 250, [\href{http://arxiv.org/abs/2012.02567}{{\tt
  arXiv:2012.02567}}].

\bibitem{Cacciari:2015jma}
M.~Cacciari, F.~A. Dreyer, A.~Karlberg, G.~P. Salam, and G.~Zanderighi, {\it
  {Fully Differential Vector-Boson-Fusion Higgs Production at
  Next-to-Next-to-Leading Order}},  {\em Phys. Rev. Lett.} {\bf 115} (2015),
  no.~8 082002, [\href{http://arxiv.org/abs/1506.02660}{{\tt
  arXiv:1506.02660}}]. [Erratum: Phys.Rev.Lett. 120, 139901 (2018)].

\bibitem{paper3}
O.~Braun-White, N.~Glover, and C.~T. Preuss, {\it {A general algorithm to build
  mixed real and virtual antenna functions for higher-order calculations}},
  \href{http://arxiv.org/abs/2307.14999}{{\tt arXiv:2307.14999}}.

\bibitem{Gehrmann:2023dxm}
T.~Gehrmann, E.~W.~N. Glover, and M.~Marcoli, {\it {The colourful antenna
  subtraction method}},  \href{http://arxiv.org/abs/2310.19757}{{\tt
  arXiv:2310.19757}}.

\bibitem{Devoto:2023rpv}
F.~Devoto, K.~Melnikov, R.~R\"ontsch, C.~Signorile-Signorile, and D.~M.
  Tagliabue, {\it {A fresh look at the nested soft-collinear subtraction
  scheme: NNLO QCD corrections to $N$-gluon final states in $q\bar{q}$
  annihilation}},  \href{http://arxiv.org/abs/2310.17598}{{\tt
  arXiv:2310.17598}}.

\bibitem{Anastasiou:2015vya}
C.~Anastasiou, C.~Duhr, F.~Dulat, F.~Herzog, and B.~Mistlberger, {\it {Higgs
  Boson Gluon-Fusion Production in QCD at Three Loops}},  {\em Phys. Rev.
  Lett.} {\bf 114} (2015) 212001, [\href{http://arxiv.org/abs/1503.06056}{{\tt
  arXiv:1503.06056}}].

\bibitem{Anastasiou:2016cez}
C.~Anastasiou, C.~Duhr, F.~Dulat, E.~Furlan, T.~Gehrmann, F.~Herzog,
  A.~Lazopoulos, and B.~Mistlberger, {\it {High precision determination of the
  gluon fusion Higgs boson cross-section at the LHC}},  {\em JHEP} {\bf 05}
  (2016) 058, [\href{http://arxiv.org/abs/1602.00695}{{\tt arXiv:1602.00695}}].

\bibitem{Mistlberger:2018etf}
B.~Mistlberger, {\it {Higgs boson production at hadron colliders at N$^{3}$LO
  in QCD}},  {\em JHEP} {\bf 05} (2018) 028,
  [\href{http://arxiv.org/abs/1802.00833}{{\tt arXiv:1802.00833}}].

\bibitem{Dreyer:2016oyx}
F.~A. Dreyer and A.~Karlberg, {\it {Vector-Boson Fusion Higgs Production at
  Three Loops in QCD}},  {\em Phys. Rev. Lett.} {\bf 117} (2016), no.~7 072001,
  [\href{http://arxiv.org/abs/1606.00840}{{\tt arXiv:1606.00840}}].

\bibitem{Duhr:2019kwi}
C.~Duhr, F.~Dulat, and B.~Mistlberger, {\it {Higgs Boson Production in
  Bottom-Quark Fusion to Third Order in the Strong Coupling}},  {\em Phys. Rev.
  Lett.} {\bf 125} (2020), no.~5 051804,
  [\href{http://arxiv.org/abs/1904.09990}{{\tt arXiv:1904.09990}}].

\bibitem{Duhr:2020kzd}
C.~Duhr, F.~Dulat, V.~Hirschi, and B.~Mistlberger, {\it {Higgs production in
  bottom quark fusion: matching the 4- and 5-flavour schemes to third order in
  the strong coupling}},  {\em JHEP} {\bf 08} (2020), no.~08 017,
  [\href{http://arxiv.org/abs/2004.04752}{{\tt arXiv:2004.04752}}].

\bibitem{Chen:2019lzz}
L.-B. Chen, H.~T. Li, H.-S. Shao, and J.~Wang, {\it {Higgs boson pair
  production via gluon fusion at N$^3$LO in QCD}},  {\em Phys. Lett. B} {\bf
  803} (2020) 135292, [\href{http://arxiv.org/abs/1909.06808}{{\tt
  arXiv:1909.06808}}].

\bibitem{Currie:2018fgr}
J.~Currie, T.~Gehrmann, E.~W.~N. Glover, A.~Huss, J.~Niehues, and A.~Vogt, {\it
  {N$^{3}$LO corrections to jet production in deep inelastic scattering using
  the Projection-to-Born method}},  {\em JHEP} {\bf 05} (2018) 209,
  [\href{http://arxiv.org/abs/1803.09973}{{\tt arXiv:1803.09973}}].

\bibitem{Dreyer:2018qbw}
F.~A. Dreyer and A.~Karlberg, {\it {Vector-Boson Fusion Higgs Pair Production
  at N$^3$LO}},  {\em Phys. Rev. D} {\bf 98} (2018), no.~11 114016,
  [\href{http://arxiv.org/abs/1811.07906}{{\tt arXiv:1811.07906}}].

\bibitem{Duhr:2020sdp}
C.~Duhr, F.~Dulat, and B.~Mistlberger, {\it {Charged current Drell-Yan
  production at N$^{3}$LO}},  {\em JHEP} {\bf 11} (2020) 143,
  [\href{http://arxiv.org/abs/2007.13313}{{\tt arXiv:2007.13313}}].

\bibitem{Duhr:2020seh}
C.~Duhr, F.~Dulat, and B.~Mistlberger, {\it {Drell-Yan Cross Section to Third
  Order in the Strong Coupling Constant}},  {\em Phys. Rev. Lett.} {\bf 125}
  (2020), no.~17 172001, [\href{http://arxiv.org/abs/2001.07717}{{\tt
  arXiv:2001.07717}}].

\bibitem{Dulat:2017prg}
F.~Dulat, B.~Mistlberger, and A.~Pelloni, {\it {Differential Higgs production
  at N$^{3}$LO beyond threshold}},  {\em JHEP} {\bf 01} (2018) 145,
  [\href{http://arxiv.org/abs/1710.03016}{{\tt arXiv:1710.03016}}].

\bibitem{Dulat:2018bfe}
F.~Dulat, B.~Mistlberger, and A.~Pelloni, {\it {Precision predictions at
  N$^3$LO for the Higgs boson rapidity distribution at the LHC}},  {\em Phys.
  Rev. D} {\bf 99} (2019), no.~3 034004,
  [\href{http://arxiv.org/abs/1810.09462}{{\tt arXiv:1810.09462}}].

\bibitem{Cieri:2018oms}
L.~Cieri, X.~Chen, T.~Gehrmann, E.~W.~N. Glover, and A.~Huss, {\it {Higgs boson
  production at the LHC using the $q_T$ subtraction formalism at N$^3$LO QCD}},
   {\em JHEP} {\bf 02} (2019) 096, [\href{http://arxiv.org/abs/1807.11501}{{\tt
  arXiv:1807.11501}}].

\bibitem{Chen:2021isd}
X.~Chen, T.~Gehrmann, E.~W.~N. Glover, A.~Huss, B.~Mistlberger, and A.~Pelloni,
  {\it {Fully Differential Higgs Boson Production to Third Order in QCD}},
  {\em Phys. Rev. Lett.} {\bf 127} (2021), no.~7 072002,
  [\href{http://arxiv.org/abs/2102.07607}{{\tt arXiv:2102.07607}}].

\bibitem{Chen:2021vtu}
X.~Chen, T.~Gehrmann, N.~Glover, A.~Huss, T.-Z. Yang, and H.~X. Zhu, {\it
  {Dilepton Rapidity Distribution in Drell-Yan Production to Third Order in
  QCD}},  {\em Phys. Rev. Lett.} {\bf 128} (2022), no.~5 052001,
  [\href{http://arxiv.org/abs/2107.09085}{{\tt arXiv:2107.09085}}].

\bibitem{Billis:2021ecs}
G.~Billis, B.~Dehnadi, M.~A. Ebert, J.~K.~L. Michel, and F.~J. Tackmann, {\it
  {Higgs pT Spectrum and Total Cross Section with Fiducial Cuts at Third
  Resummed and Fixed Order in QCD}},  {\em Phys. Rev. Lett.} {\bf 127} (2021),
  no.~7 072001, [\href{http://arxiv.org/abs/2102.08039}{{\tt
  arXiv:2102.08039}}].

\bibitem{Chen:2022cgv}
X.~Chen, T.~Gehrmann, E.~W.~N. Glover, A.~Huss, P.~Monni, E.~Re, L.~Rottoli,
  and P.~Torrielli, {\it {Third order fiducial predictions for Drell-Yan at the
  LHC}},  \href{http://arxiv.org/abs/2203.01565}{{\tt arXiv:2203.01565}}.

\bibitem{Neumann:2022lft}
T.~Neumann and J.~Campbell, {\it {Fiducial Drell-Yan production at the LHC
  improved by transverse-momentum resummation at N$^4$LL+N$^3$LO}},
  \href{http://arxiv.org/abs/2207.07056}{{\tt arXiv:2207.07056}}.

\bibitem{Camarda:2021ict}
S.~Camarda, L.~Cieri, and G.~Ferrera, {\it {Drell\textendash{}Yan lepton-pair
  production: ${q_T}$ resummation at N3LL accuracy and fiducial cross sections
  at N3LO}},  {\em Phys. Rev. D} {\bf 104} (2021), no.~11 L111503,
  [\href{http://arxiv.org/abs/2103.04974}{{\tt arXiv:2103.04974}}].

\bibitem{Chen:2022lwc}
X.~Chen, T.~Gehrmann, N.~Glover, A.~Huss, T.-Z. Yang, and H.~X. Zhu, {\it
  {Transverse Mass Distribution and Charge Asymmetry in W Boson Production to
  Third Order in QCD}},  \href{http://arxiv.org/abs/2205.11426}{{\tt
  arXiv:2205.11426}}.

\bibitem{Baglio:2022wzu}
J.~Baglio, C.~Duhr, B.~Mistlberger, and R.~Szafron, {\it {Inclusive production
  cross sections at N$^{3}$LO}},  {\em JHEP} {\bf 12} (2022) 066,
  [\href{http://arxiv.org/abs/2209.06138}{{\tt arXiv:2209.06138}}].

\bibitem{Catani:2019nqv}
S.~Catani, D.~Colferai, and A.~Torrini, {\it {Triple (and quadruple) soft-gluon
  radiation in QCD hard scattering}},  {\em JHEP} {\bf 01} (2020) 118,
  [\href{http://arxiv.org/abs/1908.01616}{{\tt arXiv:1908.01616}}].

\bibitem{DelDuca:1999iql}
V.~Del~Duca, A.~Frizzo, and F.~Maltoni, {\it {Factorization of tree QCD
  amplitudes in the high-energy limit and in the collinear limit}},  {\em Nucl.
  Phys. B} {\bf 568} (2000) 211--262,
  [\href{http://arxiv.org/abs/hep-ph/9909464}{{\tt hep-ph/9909464}}].

\bibitem{DelDuca:2019ggv}
V.~Del~Duca, C.~Duhr, R.~Haindl, A.~Lazopoulos, and M.~Michel, {\it {Tree-level
  splitting amplitudes for a quark into four collinear partons}},  {\em JHEP}
  {\bf 02} (2020) 189, [\href{http://arxiv.org/abs/1912.06425}{{\tt
  arXiv:1912.06425}}].

\bibitem{DelDuca:2020vst}
V.~Del~Duca, C.~Duhr, R.~Haindl, A.~Lazopoulos, and M.~Michel, {\it {Tree-level
  splitting amplitudes for a gluon into four collinear partons}},  {\em JHEP}
  {\bf 10} (2020) 093, [\href{http://arxiv.org/abs/2007.05345}{{\tt
  arXiv:2007.05345}}].

\bibitem{DelDuca:2022noh}
V.~Del~Duca, C.~Duhr, R.~Haindl, and Z.~Liu, {\it {Tree-level soft emission of
  a quark pair in association with a gluon}},
  \href{http://arxiv.org/abs/2206.01584}{{\tt arXiv:2206.01584}}.

\bibitem{Catani:2003vu}
S.~Catani, D.~de~Florian, and G.~Rodrigo, {\it {The Triple collinear limit of
  one loop QCD amplitudes}},  {\em Phys. Lett. B} {\bf 586} (2004) 323--331,
  [\href{http://arxiv.org/abs/hep-ph/0312067}{{\tt hep-ph/0312067}}].

\bibitem{Sborlini:2014mpa}
G.~F.~R. Sborlini, D.~de~Florian, and G.~Rodrigo, {\it {Triple collinear
  splitting functions at NLO for scattering processes with photons}},  {\em
  JHEP} {\bf 10} (2014) 161, [\href{http://arxiv.org/abs/1408.4821}{{\tt
  arXiv:1408.4821}}].

\bibitem{Badger:2015cxa}
S.~Badger, F.~Buciuni, and T.~Peraro, {\it {One-loop triple collinear splitting
  amplitudes in QCD}},  {\em JHEP} {\bf 09} (2015) 188,
  [\href{http://arxiv.org/abs/1507.05070}{{\tt arXiv:1507.05070}}].

\bibitem{Zhu:2020ftr}
Y.~J. Zhu, {\it {Double soft current at one-loop in QCD}},
  \href{http://arxiv.org/abs/2009.08919}{{\tt arXiv:2009.08919}}.

\bibitem{Catani:2021kcy}
S.~Catani and L.~Cieri, {\it {Multiple soft radiation at one-loop order and the
  emission of a soft quark\textendash{}antiquark pair}},  {\em Eur. Phys. J. C}
  {\bf 82} (2022), no.~2 97, [\href{http://arxiv.org/abs/2108.13309}{{\tt
  arXiv:2108.13309}}].

\bibitem{Czakon:2022fqi}
M.~Czakon and S.~Sapeta, {\it {Complete collection of one-loop triple-collinear
  splitting operators for dimensionally-regulated QCD}},
  \href{http://arxiv.org/abs/2204.11801}{{\tt arXiv:2204.11801}}.

\bibitem{Bern:2004cz}
Z.~Bern, L.~J. Dixon, and D.~A. Kosower, {\it {Two-loop $g\to gg$ splitting
  amplitudes in QCD}},  {\em JHEP} {\bf 08} (2004) 012,
  [\href{http://arxiv.org/abs/hep-ph/0404293}{{\tt hep-ph/0404293}}].

\bibitem{Badger:2004uk}
S.~D. Badger and E.~W.~N. Glover, {\it {Two loop splitting functions in QCD}},
  {\em JHEP} {\bf 07} (2004) 040,
  [\href{http://arxiv.org/abs/hep-ph/0405236}{{\tt hep-ph/0405236}}].

\bibitem{Duhr:2014nda}
C.~Duhr, T.~Gehrmann, and M.~Jaquier, {\it {Two-loop splitting amplitudes and
  the single-real contribution to inclusive Higgs production at N$^3$LO}},
  {\em JHEP} {\bf 02} (2015) 077, [\href{http://arxiv.org/abs/1411.3587}{{\tt
  arXiv:1411.3587}}].

\bibitem{Li:2013lsa}
Y.~Li and H.~X. Zhu, {\it {Single soft gluon emission at two loops}},  {\em
  JHEP} {\bf 11} (2013) 080, [\href{http://arxiv.org/abs/1309.4391}{{\tt
  arXiv:1309.4391}}].

\bibitem{Duhr:2013msa}
C.~Duhr and T.~Gehrmann, {\it {The two-loop soft current in dimensional
  regularization}},  {\em Phys. Lett. B} {\bf 727} (2013) 452--455,
  [\href{http://arxiv.org/abs/1309.4393}{{\tt arXiv:1309.4393}}].

\bibitem{Jakubcik:2022zdi}
P.~Jakub\v{c}\'\i{}k, M.~Marcoli, and G.~Stagnitto, {\it {The parton-level
  structure of e$^{+}$e$^{-}$ to 2 jets at N$^{3}$LO}},  {\em JHEP} {\bf 01}
  (2023) 168, [\href{http://arxiv.org/abs/2211.08446}{{\tt arXiv:2211.08446}}].

\bibitem{Chen:2023fba}
X.~Chen, P.~Jakub\v{c}\'\i{}k, M.~Marcoli, and G.~Stagnitto, {\it {The
  parton-level structure of Higgs decays to hadrons at N$^{3}$LO}},  {\em JHEP}
  {\bf 06} (2023) 185, [\href{http://arxiv.org/abs/2304.11180}{{\tt
  arXiv:2304.11180}}].

\bibitem{Chen:2023egx}
X.~Chen, P.~Jakub\v{c}\'\i{}k, M.~Marcoli, and G.~Stagnitto, {\it {Radiation
  from a gluon-gluino colour-singlet dipole at N$^3$LO}},
  \href{http://arxiv.org/abs/2310.13062}{{\tt arXiv:2310.13062}}.

\bibitem{Gehrmann-DeRidder:2005alt}
A.~Gehrmann-De~Ridder, T.~Gehrmann, and E.~W.~N. Glover, {\it {Gluon-gluon
  antenna functions from Higgs boson decay}},  {\em Phys. Lett. B} {\bf 612}
  (2005) 49--60, [\href{http://arxiv.org/abs/hep-ph/0502110}{{\tt
  hep-ph/0502110}}].

\bibitem{Gehrmann-DeRidder:2005svg}
A.~Gehrmann-De~Ridder, T.~Gehrmann, and E.~W.~N. Glover, {\it {Quark-gluon
  antenna functions from neutralino decay}},  {\em Phys. Lett. B} {\bf 612}
  (2005) 36--48, [\href{http://arxiv.org/abs/hep-ph/0501291}{{\tt
  hep-ph/0501291}}].

\bibitem{Daleo:2006xa}
A.~Daleo, T.~Gehrmann, and D.~Maitre, {\it {Antenna subtraction with hadronic
  initial states}},  {\em JHEP} {\bf 04} (2007) 016,
  [\href{http://arxiv.org/abs/hep-ph/0612257}{{\tt hep-ph/0612257}}].

\bibitem{Daleo:2009yj}
A.~Daleo, A.~Gehrmann-De~Ridder, T.~Gehrmann, and G.~Luisoni, {\it {Antenna
  subtraction at NNLO with hadronic initial states: initial-final
  configurations}},  {\em JHEP} {\bf 01} (2010) 118,
  [\href{http://arxiv.org/abs/0912.0374}{{\tt arXiv:0912.0374}}].

\bibitem{Pires:2010jv}
J.~Pires and E.~W.~N. Glover, {\it {Double real radiation corrections to gluon
  scattering at NNLO}},  {\em Nucl. Phys. B Proc. Suppl.} {\bf 205-206} (2010)
  176--181, [\href{http://arxiv.org/abs/1006.1849}{{\tt arXiv:1006.1849}}].

\bibitem{Boughezal:2010mc}
R.~Boughezal, A.~Gehrmann-De~Ridder, and M.~Ritzmann, {\it {Antenna subtraction
  at NNLO with hadronic initial states: double real radiation for
  initial-initial configurations with two quark flavours}},  {\em JHEP} {\bf
  02} (2011) 098, [\href{http://arxiv.org/abs/1011.6631}{{\tt
  arXiv:1011.6631}}].

\bibitem{Gehrmann:2011wi}
T.~Gehrmann and P.~F. Monni, {\it {Antenna subtraction at NNLO with hadronic
  initial states: real-virtual initial-initial configurations}},  {\em JHEP}
  {\bf 12} (2011) 049, [\href{http://arxiv.org/abs/1107.4037}{{\tt
  arXiv:1107.4037}}].

\bibitem{Gehrmann-DeRidder:2012too}
A.~Gehrmann-De~Ridder, T.~Gehrmann, and M.~Ritzmann, {\it {Antenna subtraction
  at NNLO with hadronic initial states: double real initial-initial
  configurations}},  {\em JHEP} {\bf 10} (2012) 047,
  [\href{http://arxiv.org/abs/1207.5779}{{\tt arXiv:1207.5779}}].

\bibitem{Gehrmann-DeRidder:2009lyc}
A.~Gehrmann-De~Ridder and M.~Ritzmann, {\it {NLO Antenna Subtraction with
  Massive Fermions}},  {\em JHEP} {\bf 07} (2009) 041,
  [\href{http://arxiv.org/abs/0904.3297}{{\tt arXiv:0904.3297}}].

\bibitem{Abelof:2011ap}
G.~Abelof and A.~Gehrmann-De~Ridder, {\it {Double real radiation corrections to
  $t\bar{t}$ production at the LHC: the all-fermion processes}},  {\em JHEP}
  {\bf 04} (2012) 076, [\href{http://arxiv.org/abs/1112.4736}{{\tt
  arXiv:1112.4736}}].

\bibitem{Bernreuther:2011jt}
W.~Bernreuther, C.~Bogner, and O.~Dekkers, {\it {The real radiation antenna
  function for $S \to Q {\bar Q} q {\bar q}$ at NNLO QCD}},  {\em JHEP} {\bf
  06} (2011) 032, [\href{http://arxiv.org/abs/1105.0530}{{\tt
  arXiv:1105.0530}}].

\bibitem{Abelof:2011jv}
G.~Abelof and A.~Gehrmann-De~Ridder, {\it {Antenna subtraction for the
  production of heavy particles at hadron colliders}},  {\em JHEP} {\bf 04}
  (2011) 063, [\href{http://arxiv.org/abs/1102.2443}{{\tt arXiv:1102.2443}}].

\bibitem{Abelof:2012bga}
G.~Abelof and A.~Gehrmann-De~Ridder, {\it {Double real radiation corrections to
  top-antitop production at the LHC}},  {\em PoS} {\bf LL2012} (2012) 061.

\bibitem{Abelof:2012rv}
G.~Abelof and A.~Gehrmann-De~Ridder, {\it {Double real radiation corrections to
  $t\bar{t}$ production at the LHC: the $gg\rightarrow t\bar{t}q\bar{q}$
  channel}},  {\em JHEP} {\bf 11} (2012) 074,
  [\href{http://arxiv.org/abs/1207.6546}{{\tt arXiv:1207.6546}}].

\bibitem{Bernreuther:2013uma}
W.~Bernreuther, C.~Bogner, and O.~Dekkers, {\it {The real radiation antenna
  functions for $S\rightarrow Q\bar{Q}gg$ at NNLO QCD}},  {\em JHEP} {\bf 10}
  (2013) 161, [\href{http://arxiv.org/abs/1309.6887}{{\tt arXiv:1309.6887}}].

\bibitem{Dekkers:2014hna}
O.~Dekkers and W.~Bernreuther, {\it {The real-virtual antenna functions for $S
  \to Q\bar{Q} X$ at NNLO QCD}},  {\em Phys. Lett. B} {\bf 738} (2014)
  325--333, [\href{http://arxiv.org/abs/1409.3124}{{\tt arXiv:1409.3124}}].

\bibitem{Gustafson:1987rq}
G.~Gustafson and U.~Pettersson, {\it {Dipole Formulation of QCD Cascades}},
  {\em Nucl. Phys. B} {\bf 306} (1988) 746--758.

\bibitem{Lonnblad:1992tz}
L.~Lonnblad, {\it {ARIADNE version 4: A Program for simulation of QCD cascades
  implementing the color dipole model}},  {\em Comput. Phys. Commun.} {\bf 71}
  (1992) 15--31.

\bibitem{Giele:2007di}
W.~T. Giele, D.~A. Kosower, and P.~Z. Skands, {\it {A simple shower and
  matching algorithm}},  {\em Phys. Rev. D} {\bf 78} (2008) 014026,
  [\href{http://arxiv.org/abs/0707.3652}{{\tt arXiv:0707.3652}}].

\bibitem{Giele:2011cb}
W.~T. Giele, D.~A. Kosower, and P.~Z. Skands, {\it {Higher-Order Corrections to
  Timelike Jets}},  {\em Phys. Rev. D} {\bf 84} (2011) 054003,
  [\href{http://arxiv.org/abs/1102.2126}{{\tt arXiv:1102.2126}}].

\bibitem{Fischer:2016vfv}
N.~Fischer, S.~Prestel, M.~Ritzmann, and P.~Skands, {\it {Vincia for Hadron
  Colliders}},  {\em Eur. Phys. J. C} {\bf 76} (2016), no.~11 589,
  [\href{http://arxiv.org/abs/1605.06142}{{\tt arXiv:1605.06142}}].

\bibitem{Brooks:2020upa}
H.~Brooks, C.~T. Preuss, and P.~Skands, {\it {Sector Showers for Hadron
  Collisions}},  {\em JHEP} {\bf 07} (2020) 032,
  [\href{http://arxiv.org/abs/2003.00702}{{\tt arXiv:2003.00702}}].

\bibitem{Li:2016yez}
H.~T. Li and P.~Skands, {\it {A framework for second-order parton showers}},
  {\em Phys. Lett. B} {\bf 771} (2017) 59--66,
  [\href{http://arxiv.org/abs/1611.00013}{{\tt arXiv:1611.00013}}].

\bibitem{Campbell:2021svd}
J.~M. Campbell, S.~H\"oche, H.~T. Li, C.~T. Preuss, and P.~Skands, {\it
  {Towards NNLO+PS matching with sector showers}},  {\em Phys. Lett. B} {\bf
  836} (2023) 137614, [\href{http://arxiv.org/abs/2108.07133}{{\tt
  arXiv:2108.07133}}].

\bibitem{Altarelli:1977zs}
G.~Altarelli and G.~Parisi, {\it {Asymptotic Freedom in Parton Language}},
  {\em Nucl. Phys. B} {\bf 126} (1977) 298--318.

\bibitem{Dokshitzer:1977sg}
Y.~L. Dokshitzer, {\it {Calculation of the Structure Functions for Deep
  Inelastic Scattering and $e^+ e^-$ Annihilation by Perturbation Theory in
  Quantum Chromodynamics.}},  {\em Sov. Phys. JETP} {\bf 46} (1977) 641--653.

\bibitem{Weinzierl:2006wi}
S.~Weinzierl, {\it {Status of jet cross sections to NNLO}},  {\em Nucl. Phys. B
  Proc. Suppl.} {\bf 160} (2006) 126--130,
  [\href{http://arxiv.org/abs/hep-ph/0606301}{{\tt hep-ph/0606301}}].

\bibitem{ALTARELLI1979461}
G.~Altarelli, R.~Ellis, and G.~Martinelli, {\it Large perturbative corrections
  to the drell-yan process in qcd},  {\em Nuclear Physics B} {\bf 157} (1979),
  no.~3 461--497.

\bibitem{Catani:1998nv}
S.~Catani and M.~Grazzini, {\it {Collinear factorization and splitting
  functions for next-to-next-to-leading order QCD calculations}},  {\em Phys.
  Lett. B} {\bf 446} (1999) 143--152,
  [\href{http://arxiv.org/abs/hep-ph/9810389}{{\tt hep-ph/9810389}}].

\end{thebibliography}\endgroup
\end{document}